%% file: main.tex
\documentclass[aps, prx, reprint, superscriptaddress]{revtex4-2}

\usepackage{graphicx}
\usepackage{dcolumn}
\usepackage{bm}
\usepackage{xcolor}
\usepackage{amssymb}
\usepackage{amsmath}
\usepackage{physics}
\usepackage{mathtools}
\usepackage{titlesec}
\usepackage[force]{feynmp-auto}
\usepackage{epsfig}
\usepackage{lineno}
\usepackage{appendix}
\usepackage{lipsum}
\usepackage{tabularx}
\usepackage{booktabs}

\def\ik{\textbf{\textit{k}}}

\def\ip{\textbf{\textit{p}}}
\def\iq{\textbf{\textit{q}}}

\def\iv{{i\nu}}
\def\io{{i\omega}}
\def\iw{{i\omega}}
\def\iO{{i\Omega}}

\def\Soneloop{\Sigma^{\text{1loop}}}

\newcommand{\ive}[1]{\textbf{\textit{#1}}}

\fmfset{arrow_len}{6}
\fmfset{arrow_ang}{20}

\begin{document}
  \title{{Quantum fluctuations and the emergence of in-gap Higgs mode in superconductors}}

  \author{S. Tian}
  \email[]{s.tian@fkf.mpg.de}
  \affiliation{Max Planck Institute for Solid State Research, Heisenbergstraße 1, D-70569 Stuttgart, Germany}

  \author{N. Tsuji}
  \affiliation{Department of Physics, The University of Tokyo, Hongo, Tokyo 113-8656, Japan}
  \affiliation{RIKEN Center for Emergent Matter Science (CEMS), Wako 351-0198, Japan}
  \affiliation{Trans-Scale Quantum Science Institute, The University of Tokyo, Hongo, Tokyo 113-8656, Japan}

  \author{D. Manske}
  \affiliation{Max Planck Institute for Solid State Research, Heisenbergstraße 1, D-70569 Stuttgart, Germany}

  \date{\today}

  \begin{abstract}
    We extend the well-established action of the Higgs mode in $s$-wave
    superconductors to include quantum fluctuations (QFs). We find that already one-loop quantum corrections to the Higgs propagator shift its eigenfrequency below the
    superconducting energy gap $2\Delta$.
    {Consequently, the Higgs mode appears as an undamped pole below the quasiparticle continuum, leading to drastically sharper experimental signatures.}
    We {demonstrate} this by calculating two characteristic fingerprints of the {Higgs mode}, namely {in} Third
    Harmonic Generation (THG) and {inelastic Raman scattering signals}. {More generally, gaps measured in $s$-wave superconductors with different experimental techniques (such as scanning tunneling microscope and Raman scattering) may be different due to fluctuation corrections.}{ Since already arbitrarily weak QFs lead to the shift and to the new pole, our results shed some light on other amplitude modes even for systems with weak QFs, including charge density waves, (anti-) ferromagnets, or cold atom fermionic condensates.}
  \end{abstract}

  \maketitle

  \section{Introduction}
  In general, a superconductor is characterised by the spontaneously broken $U(1)$ symmetry, where a complex order parameter $\Delta$ emerges below the critical temperature $T_c$. This is accompanied by two collective dynamics of $\Delta$: the Higgs mode and the Goldstone mode (see \cite{pekker_review,shimano_2020} for reviews) corresponding to amplitude and phase fluctuations of $\Delta$, respectively. {In superconductors}, the Goldstone mode is pushed to the plasma frequency due to the Anderson-Higgs mechanism \cite{Anderson_RPA,Nambu}, but the Higgs mode remains unaltered \cite{Littlewood_Verma}. As a coherent oscillation of the condensate, the Higgs mode can serve as a fingerprint of the underlying ground state \cite{Lukas_2020_classification,barlasAmplitudeHiggsModes2013,neri2025collectivemodespectroscopytimereversal}. This {new field of Higgs spectroscopy} inspired numerous experiments including but not limited to: pump-probe spectroscopy \cite{Matsunaga_2013, Katsumi_2018,katsumi_2020,luo_2D,vaswaniLightQuantumControl_2D}, multicycle terahertz third harmonic generation (THG) \cite{Matsunaga_2014, Chu_2020,liwen}, direct current assistant THG generation \cite{Nakamura_2019}, and non-equilibrium antistokes Raman spectroscopy (NEARS) \cite{NEARS}.

Nevertheless, experimental identification of the Higgs mode in superconductors has a twofold difficulty: Firstly, the Higgs mode couples only quadratically to light, yielding weak spectral signatures that are often overshadowed by {broken Cooper pairs, e.g. by }quasiparticle (QP) excitations \cite{Lara_cooperpairVSHiggs}. This can be circumvented with the help of extrinsic effects such as impurities, which enhance the Higgs mode response drastically via allowing paramagnetic coupling to light \cite{murotani_impurity,silaev_impurity,tsuji_impurity,seibold_impurity,rafael_impurity}. Secondly, the Higgs mode lacks a well-defined peak, {since the} corresponding spectral function lacks a pole.{ In $s$-wave superconductors, weak-coupling theories predict a degeneracy between the Higgs mode eigenfrequency ($\omega_H$) and pair excitation gap ($E_g=2\Delta$), leading to a damped weak square-root divergence \cite{ceaNonrelativisticDynamicsAmplitude2015}. In the {unconventional case such as $d$-wave superconductors}, nodal QPs will further smear out the divergence into a broad peak \cite{Lukas_2020}}.  {Thus, to summarize, the lack of a sharp pole makes the Higgs mode inherently difficult to observe.}

Recent studies beyond the BCS approximation have shown that the Higgs mode in $s$-wave superconductors can exhibit a spectral pole inside the gap (therefore $\omega_H \neq E_g$) under special conditions. These include helical superconductors coupling to an LC circuit \cite{luReducingFrequencyHiggs2023}, adding magnetic impurities \cite{Li_ingap_magnetic_impurity}, and modifying the pairing interaction \cite{althuser_modified_GUT_ingaphiggs}. This was also observed recently in cold-atom experiments \cite{cabreraHybridizationAmplitudeMode2025}. Notably, in the strong coupling regime, several theoretical works incorporating beyond-mean-field effects have reported an in-gap Higgs mode: strong coupling corrections and polarization of the fermionic vacuum of the spin-triplet $p$-wave superfluid $^3$He \cite{saulsNambuFermionbosonRelations2017}, a dynamical mean field theory study of the Holstein model \cite{park_holstein_BCSBEC}, and two investigations of the Hubbard model using the two-particle self-consistent approach \cite{lorenzo_two_particle_self_consistent} and time-dependent Gutzwiller approach \cite{lorenzanaLongLivedHiggsModes2024a}.

{While the appearance of the in-gap Higgs mode seems to be a strong-coupling result, the lack of a discontinuous transition from the weak to strong coupling in the BCS-BEC crossover implies similar phenomena near the weak-coupling limit.} {We will build on top of this intuition, and aim for a more realistic description of the Higgs mode that extends our current understanding of its properties. We found that quantum fluctuations perturbatively push the Higgs mode inside the gap. Our findings imply that $\omega_H$ and $E_g$ decouple for an arbitrarily weak attractive potential, and the simple relation $\omega_H = E_g=2\Delta$ is only a manifestation of the mean field ansatz. This is achieved by reinterpreting the Luttinger-Ward functional as a functional of the Higgs mode and expanding to two-loop order. {Since the degeneracy $\omega_H = E_g = 2\Delta$ is lifted}, a Higgs mode with power-law divergence appears inside the gap. Including QFs thus constitutes a \textit{qualitatively different} theory, {accompanied by a significantly enhanced Higgs mode response}. }

This work is organized as follows. In Sec.~(\ref{sec:motivation}) we first motivate the search for a Higgs mode pole beyond the Random Phase approximation. In Sec.~(\ref{sec:effective_theory}) we present our theoretical framework and clarify some notations. In Sec.~(\ref{sec:results}), we solve the Dyson equation for the Higgs mode numerically with the self-energy evaluated at one-loop order, leading to a well-defined Higgs peak inside the pair-breaking gap. Later in Sec.~(\ref{sec:response_functions}), we compute the Raman and THG response functions of the Higgs mode and comment on the potential observability of the in-gap mode in materials. We will discuss the outlook of this theory in Sec.~(\ref{sec:conclusion}).

  
  \section{The Higgs propagator and a square root divergence} \label{sec:motivation}
  Before turning to the formal mathematics, we believe it is useful to present an intuitive picture of the mechanism behind the appearance of a Higgs mode pole. Amplitude fluctuations of the superconducting order parameter \(\Delta(t)\) in superconductors are commonly referred to as the  \emph{Higgs mode} (see \cite{shimano_2020} and references therein) owing to their analogy to relativistic field theories: both are effective theories of complex scalar fields coupled to gauge fields. It is well established that these fluctuations decay even in the absence of particle collisions \cite{VolkovKogan1973,yuzbashyanDynamicalVanishingOrder2006a,Cea_cdw}:
    \[
\delta \Delta (t) \propto \frac{\cos(2\Delta_{\infty }t)}{\sqrt{ t }}\, ,
\]
    where \(\delta \Delta(t) = \Delta(t) - \Delta_{\infty} \) denotes fluctuation about the long-time limit \(\Delta_{\infty} = \Delta(t \rightarrow \infty)\). Within the so-called pseudospin formalism \cite{tsujiTheoryAndersonPseudospin2015a,Lukas_2020}, such a decay stems from the dephasing among pseudospin precessions \cite{yuzbashyanDynamicalVanishingOrder2006a,Cea_cdw}: pseudospins at similar momenta \(\ik\) precess at slightly different frequencies, which become incoherent in the long-time limit. From a field theoretical perspective, the \({1}/{\sqrt{ t }}\) decay reflects a square-root divergence of the correlator \(\expval{\delta\Delta\delta\Delta}\) in the long wave limit.
  At the Random Phase approximation (RPA) level, the Higgs propagator \(\expval{\delta\Delta\delta\Delta}\) in the $\iq$ = 0 limit can be written as \cite{Varma_littlewood_nbse2,kamataniOpticalResponseLeggett2022a}:
\begin{eqnarray} 
  H_{0,\iq=0}^{-1}(i\omega)
  &=& \sum_{\ik} \frac{4\Delta^2 - (i\omega)^2}
  {E_\ik \left(4E_\ik^2 - (i\omega)^2\right)} \nonumber \\
  &=& N_F \int_{-\infty}^{\infty} d\xi
  \frac{4\Delta^2 - (i\omega)^2}
  {\sqrt{\Delta^2+\xi^2}\left(4\Delta^2+4\xi^2-(i\omega)^2\right)} \nonumber \\
  &=& N_F \left(4\Delta^2 - (i\omega)^2\right) F(i\omega), \label{eqn:HiggsPropagator}
\end{eqnarray}
with
\begin{equation}
  F(i\omega) =
  \frac{2\arcsin\!\left(i\omega / 2\Delta\right)}
  {i\omega\sqrt{4\Delta^2-(i\omega)^2}} \, .
\end{equation}
In the second step, the momentum sum was replaced by an integral over the dispersion
$\xi$ and the density of states at the Fermi surface $N_F$ {per spin} was introduced. Integration boundaries were pushed to infinity, as the integration converges without the need for a Debye cutoff. Near \(i\omega = 2\Delta\), the inverse Higgs propagator vanishes with a branch point \(H_{0}^{-1} \propto \sqrt{ 4\Delta^{2} -(\iw)^{2}}\). Importantly, it is the function \(F(\iw)\)  that ``smears out'' the proper singularity of \(H_{0}\). 

Note that the same function \(F(\iw) \) appears in the unscreened density-density correlator \cite{Lara_cooperpairVSHiggs}:
\begin{eqnarray}
    \expval{\rho \rho} &=& \sum_{\ik} \frac{4\Delta^{2}}{E_\ik(4E_\ik^2 - (\iw)^2)}  \nonumber\\  
    &=& 4\Delta^{2} F(\iw) \, .
\end{eqnarray}

Thus, the Higgs propagator diverges like a square root because the true pole \(4\Delta^{2} - (\iw)^2\) locates \emph{exactly} at the edge of the quasiparticle continuum. The resulting excitation is then weighted by a density-density correlation factor, replacing the Lorentzian singularity with a square-root one.

However, the Higgs propagator in Eq.~\eqref{eqn:HiggsPropagator} is a RPA-level result. Upon moving beyond RPA by including quantum fluctuations (QFs), we must compute the \emph{dressed Higgs propagator} from the long-wave-limit (\(\iq=0\)) Dyson equation:
\begin{equation}
H^{-1}(\iw) = H_{0}^{-1}(\iw) - \Sigma(\iw) \, .
\end{equation}

\begin{figure}
    \centering
    \includegraphics[width=1\linewidth]{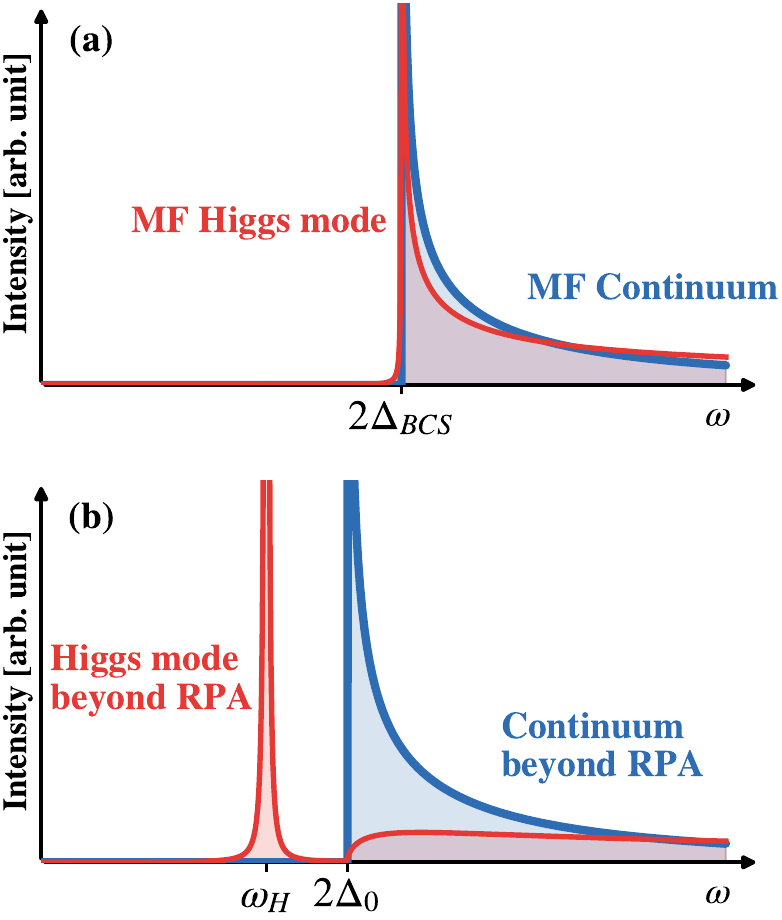}
    \caption{{Schematic view for the spectral functions of the Higgs mode and pair breaking, respectively. (a). Conventional picture of the mean field (MF) results: the Higgs mode appears exactly at the edge of the quasiparticle continuum, i.e., at $\omega = 2\Delta_{BCS}$. (b). The inclusion of quantum fluctuations (QFs) reduces the  BCS theory gap $2\Delta_{BCS}$ down to $2\Delta_0$. Additionally, the Higgs mode is pushed inside of $2\Delta_0$, revealing its Lorentzian pole.}}
    \label{fig:schematic_pic}
\end{figure}

Here \(\Sigma\), the self-energy, encodes interactions between the Higgs mode and the underlying superconducting quasiparticles, including intermediate pair-breaking processes. We therefore conjecture that, near the pair breaking energy $\iw \sim 2\Delta$,  the self-energy \(\Sigma\) is proportional to the density-density correlator, namely: 
\begin{equation}\label{eqn:PossibleEquality}
\Sigma(\iw) \stackrel{?}{=} C N_{F}F(\iw)\,  \quad \text{For \(\iw \rightarrow 2\Delta \)}. 
\end{equation}
with \(C\) being some constant of proportionality. Under this assumption, the dressed Higgs mode takes the form:
\begin{equation}\label{eqn:conjecture}
    H^{-1}(\iw) \stackrel{?}{=} N_{F}(4\Delta^{2} - C - (\iw)^2) F(\iw)\, . 
\end{equation}

Physically, this equality implies that QFs push the Higgs mode pole away from the quasiparticle continuum. We present a schematic illustration of this process in Fig.~(\ref{fig:schematic_pic}). It remains to prove the validity of Eq.~(\ref{eqn:PossibleEquality}) and show that \(C>0\); this will be established in the remaining of this paper.

\section{Effective theory for superconductors}\label{sec:effective_theory}
  We will first introduce our notations and model. Throughout this work, we stay
  in the zero temperature limit and set $\hbar = 1$. Consider the attractive Hubbard model:
  \begin{eqnarray}
    \label{eqn:hubbard_model} \mathcal{H} &=& \sum_{ij \sigma} (t_{ij} -\mu\delta_{ij} ) c^\dagger_{i \sigma}
    c_{j \sigma} -V\sum_{i} c^\dagger_{i\uparrow}
    c^\dagger_{i \downarrow} c_{i \downarrow} c_{i \uparrow}  \, ,
  \end{eqnarray}
  where the underlying lattice can be arbitrary. $V>0$ is the point-range
  attraction, $c_{i\sigma}$ $(c_{i\sigma}^{\dagger})$ is the fermionic creation (annihilation)
  operator at site $i$ with spin \(\sigma\). The hopping amplitude is $t_{ij}$, and $\mu$ is the chemical potential.
  To reveal the collective mode dynamics, a standard treatment is to perform the
  Hubbard-Stratonovich (HS) transformation on the Hamiltonian Eq.~(\ref{eqn:hubbard_model})
  (see, for example, \cite{Lara_cooperpairVSHiggs,Schwarz_2021,rafael_impurity}
  and references therein) and integrate out the fermionic degrees of freedom. Here,
  we use a slightly modified approach. Using Nambu spinors \cite{Nambu}
  $\Psi_{i}^{\dagger}= (c^{\dagger}_{i\uparrow}, c_{i\downarrow})$, we decompose
  the four-point interaction term in Eq.~\eqref{eqn:hubbard_model} by introducing
  a complex scalar field $\Delta$:
  \begin{eqnarray}
    &&S[\Psi,\Psi^\dagger,\Delta,\Delta^\dagger] \nonumber \\
    &=& \int_{-\infty}^{\infty} d\tau \, \sum_{ij}\Psi^\dagger_i(\tau)(\partial_\tau\delta_{ij}
    +t_{ij}\sigma_3-\mu \delta_{ij})\Psi_j(\tau) \nonumber \\
    &+& \frac{1}{2}\Delta^\dagger_i(\tau)\frac{2\delta_{ij}}{V} \Delta_j(\tau) +
    \sum_i\Psi^\dagger_i(\tau)
    \begin{pmatrix}
      0                          & \Delta_{i}(\tau) \\
      \Delta^{\dagger}_{i}(\tau) & 0
    \end{pmatrix}
    \Psi_i(\tau)  \, . \nonumber\\\label{eqn:yukawa_action2} 
  \end{eqnarray}

  The above action can be interpreted as a Yukawa-type theory, where a complex
  bosonic field $\Delta$ interacts with a fermionic field $\Psi$ via a two-Fermion
  one-Boson vertex. We will not integrate out the Fermions, as doing so introduces
  terms to infinite orders of the field $\Delta$; formulating a consistent
  interacting theory with an infinite set of interactions will be arduous.

  We are interested in the condensed phase of the model Eq.~\eqref{eqn:yukawa_action2},
  where $\Delta$ acquires a finite expectation value in the ground state: $\Delta
  _{i}= (\Delta_{0}+ h_{i})e^{i\theta_{i}}$. In superconductors, the dynamics of
  $\theta$ is gapped due to the long-range Coulomb repulsion \cite{Anderson_RPA,Nambu}
  whose energy scale is independent of superconductivity. We therefore ignore
  the phase dynamics by retaining only the amplitude fluctuations of $\Delta$ \footnote{This is strictly true in 3 dimensions, where the phase mode becomes the plasmon with a pole at much higher energy scales $\Omega_{P}=\frac{4\pi e^{2}\rho}{m}$ for particles with $e$, mass $m$, and density $\rho$ \cite{lara_2008_low_energy_phase_mode_only_action}. In 2 dimensions, the plasmon is gapless, but the dynamics are nevertheless at a much higher energy scale \cite{sunCollectiveModesTerahertz2020}.}. Expanding {around} the vacuum
  expectation of $\Delta$:
  \begin{equation}\label{eqn:Expansion}
    \Delta_{i}(\tau) = \Delta_{0}+ h_{i}(\tau) \, ,
  \end{equation}
  with a homogeneous \(s\)-wave superconducting gap $\Delta_{0}$. This gives rise to an
  effective action:
  \begin{eqnarray}
    S[\Psi,\Psi^{\dagger},h] &= &\int d\tau \sum_{ij}\Psi^\dagger_iG^{-1}_{0,ij}\Psi_j
    + \frac{1}{2}\frac{2\delta_{ij}}{V}h_ih_j \nonumber\\
    &+& \sum_ih_i\Psi^\dagger_i\sigma_1\Psi_i + \sum_i 2\Delta_0 h_i /V \, .
    \label{eqn:yukawa_action}
  \end{eqnarray}

  We use the Pauli matrices $\sigma_{a}(a = 1,2,3)$ to simplify our notation. Dependencies
  of the fields on the imaginary time $\tau$ are also made implicit for clarity. A set of Feynman rules arises naturally from this theory, which we present in momentum space:

  \input{feyn_macro}

  \subsection{The Luttinger-Ward functional}
  To construct a self-consistent theory, we use the Luttinger-Ward (LW) functional
  $\Phi[G,H]$ \cite{luttingerGroundStateEnergyManyFermion1960}, defined as the sum over all
  closed, two-particle-irreducible skeleton diagrams in the perturbative approach
  \cite{benlagraLuttingerWardFunctionalApproach2011}. We present the LW functional
  considered in this work in Eq.\eqref{eqn:LWfunctional}. The double-arrowed
  line represents the \emph{dressed} Nambu quasiparticle propagator $G_{\ik}(\iv)$. The
  solid-red-line represents the \emph{dressed} $h_{i}$ field propagator
  $H_{\iq}(\iw)$, which we refer to as the \emph{Higgs propagator} throughout
  this work. As proven by Luttinger and Ward \cite{luttingerGroundStateEnergyManyFermion1960},
  {physical observables obtained from this approximation will respect conservation laws.}

  \begin{widetext}
    \input{LW_macro}
  \end{widetext}

  The functional $\Phi$ is, in principle, an infinite series; however higher order
  diagrams are effectively suppressed. Let $\ik$ and $\iq$ denote the momenta associated with the Nambu propagator $G_{\ik}(\iv)$ and the Higgs propagator
$H_{\iq}(\iw)$, respectively. {Unlike the electron momentum $\ik$, the collective mode momentum $\iq$ should be bounded by the coherence length $\abs{\iq}<\xi_0^{-1}$ of the corresponding order. In the case of superconductors,  $\xi_0=\frac{v_{F}}{\sqrt{12}\Delta_{0}}$ \cite{shimano_2020,Varma_littlewood_nbse2,lara_2008_low_energy_phase_mode_only_action} with \(v_{F}\) denoting the Fermi velocity. Physically, we expect the collective dynamics of Cooper pairs to decay exponentially below the length scale $\xi_0$, which measures the average size of a Cooper pair \cite{lara_2008_low_energy_phase_mode_only_action}. This naturally gives rise to the ultraviolet cutoff. Since we consider only small $\iq$, it is justified to linearize the electronic dispersion using $\xi_{\ik\pm\iq} \approx \xi_\ik \pm \ive{v}_F\cdot \iq$. Details can be found in the supplementary material \cite{SM}. We therefore impose different momentum cutoffs for $H$ and $G$:}
  \begin{eqnarray}
    \xi_\ik &\in& [-\Lambda,\Lambda]\\
    \abs{\iq}&\in& [0,\xi_0^{-1}]
  \end{eqnarray}
  where $\Lambda$ is {the energy cutoff for \(\xi_{\ik}\)}. In the sum over electron momentum, we simplify the integral using a constant density of state:
  \begin{equation}
      \sum_\ik \to N_F \int_{-\Lambda}^{\Lambda} d\xi
  \end{equation}
  and the integration boundaries are pushed to infinity whenever the integral over $\xi$ converges. Assuming isotropy in $\iq$, the boundary on \(\iq\) allow us to isolate the \(\xi_{0}\) dependence:
  \color{black}
  \begin{eqnarray}\label{eqn:MagnitudeOfIntegral}
          \int \frac{{d^d \iq}}{(2\pi)^d} \ f(q)  &=& d K_d \int_0^{\xi_0^{-1}} q^{d-1} f(q) dq \nonumber\\
          &=& \frac{d K_d}{\xi_0^d} \int_0^1 u^{d-1} {f(\xi_0^{-1}u)} du  \nonumber \\
          &=& \alpha N_F \times d\int_0^1 u^{d-1} {f(\xi_0^{-1}u)} du \, .
  \end{eqnarray}

Here
\begin{eqnarray}\label{eqn:alpha_definition}
    \alpha &=& \frac{ K_d}{N_F \xi_0^d}\, , 
\end{eqnarray}
and \(K_{d}\) denotes volume of unit sphere divided by \((2\pi)^d\), and hence $dK_d$ gives the solid angle in $d$-dimension
{$2K_2 = 1/2\pi$ and $3K_3 = 1/2\pi^2$}. The integral in final line of Eq.~\eqref{eqn:MagnitudeOfIntegral} is of size unity for sufficiently well-behaved function \(f\), and the prefactor \(\alpha\) determines the overall magnitude of the momentum integral. 

This parameter $\alpha$ has the unit of energy, and is small compared to $\Delta_0$ in the weak-coupling limit. To see this, consider the weak coupling limit \(\Delta_{0} \ll \Lambda \ll E_{F}\), where \(E_{F}\) denotes the Fermi energy. In this regime, the electronic dispersion $\xi_\ik$ near the Fermi level is then well approximated by a parabolic band with  \(E_{F} = \frac{1}{2}v_{F} k_{F}\) and \(N_{F}= d K_{d} k_{F}^{d-1}/v_{F}\). Using these relations, we obtain:
\begin{equation}\label{eqn:alpha_delta_relation}
        \alpha/\Delta_0 = \left(\frac{\sqrt{12}^d}{2^{d-1} d}\right) \left(\frac{\Delta_0}{E_F}\right)^{d-1} \, , 
\end{equation}
where the \emph{Ginzburg parameter} 
$
G \sim  {\left( {\Delta_{0}}/{E_{F}} \right)    }^{d-1}
$
appears explicitly. Define:

\begin{equation}
    \delta = \frac{\Delta_{0}}{E_{F}}.
\end{equation}

 Each integral over the collective mode momenta therefore contributes a factor of \(\delta^{d-1}\). Hence, $\alpha/\Delta_0$ is a formal small parameter of {the expansion} {and our theory is self-validating in the sense that the QF corrections vanish in the weak-coupling limit $\delta \to 0$. Diagrammatically, counting orders of $\alpha/\Delta_0$ is equivalent to counting loop orders in collective mode momentum}.  Our truncation of \(\Phi\) (Eq.~(\ref{eqn:LWfunctional})) is therefore controlled up to order \(\mathcal{O}((\alpha/\Delta_{0})^2)\). We reserve the terminology \emph{n-loop order} to refer to terms explicitly $n$-loop in the collective mode momentum (thus to order $\mathcal{O}((\alpha/\Delta_0)^n)$).

\subsection{Self energies and Dyson equations}
So far, we have established that our Luttinger-Ward functional up to two-loop order. In this section we will use Eq.~(\ref{eqn:LWfunctional}) to construct self energies of \(G_{\ik}\) and \(H_{\iq}\). The Dyson equations for the propagators are:
\begin{eqnarray} 
    G_\ik^{-1}(\iv)  &=&  G^{-1}_{0,\ik}(\iv) - \Pi[G,H]  \label{eqn:dyson_G}\\
    H^{-1}_\iq (\io) &=& \frac{2}{V} - \Sigma[G,H]\, ,  \label{eqn:dyson_H}
\end{eqnarray}
where the self-energies are related to the LW functional via \footnote{Note that the Hartree term (the first two terms in \(\Phi\)) do not contribute to \(\Sigma\) and have to be excluded. Since these diagrams are not two-particle-irreducible for \(H\). A minus sign is needed for the {fermionic} self-energy $\Pi$ as we broke a closed loop. A factor of 2 is needed for $\Sigma$ since scalar boson propagators are symmetric with respect to switching starting and end points.}:
\begin{equation}
    \frac{\delta\Phi}{\delta G} = -\Pi; \quad\quad 2\frac{\delta\Phi|_{\text{no Hartree}}}{\delta H} = \Sigma\, , 
\end{equation}
which give rise to the following diagrams:

\begin{widetext}
\begin{equation}\label{eqn:diagramDysonG}
    \input{Dyson_G}
\end{equation}
\begin{equation}\label{eqn:diagramDysonH}
    \input{Dyson_H}
\end{equation}
\end{widetext}

An important note must be made. The Nambu quasiparticle self-energy \(\Pi\) also gives a modified superconducting gap equation. Consider the first two terms in \(\Pi\): these two must cancel, otherwise we obtain corrections to the off-diagonal component of \(G\) to zeroth loop order, implying that we are not expanding from the correct ground state value of \(\Delta_{0}\):
\newline
\newline
\input{Gap_equation_Full}

The simplest form of this gap equation is when we replace \(G_{\ik} \rightarrow G_{0,\ik}\): 

\begin{equation}\label{eqn:diagrammatic_gap_equation}
\begin{fmffile}{gap_equation}
-\frac{2\Delta_0}{V} = \quad \parbox{40pt}{
\begin{fmfgraph*}(40,40)
\fmfset{arrow_len}{7}
    \fmfleft{i}
    \fmfright{j}
    \fmf{fermion,left=1}{i,j,i}
    \fmfdot{i}
    \fmfv{l.a=0,l=$\sigma_1$}{i}
\end{fmfgraph*}}
\end{fmffile} \, {,} 
\end{equation} 
which reduces promptly into:
\begin{equation}\label{eqn:algebraic_gap_equation}
    \frac{2}{V} = \sum_{\ik} \frac{1}{\sqrt{ \xi_{\ik}^{2} + \Delta_{0}^{2} }}\, {,}
\end{equation}
the Bardeen–Cooper–Schrieffer (BCS) gap equation at zero temperature. Adding QF corrections will therefore modify \(\Delta_0\) accordingly.
 
The equations {Eq.~(\ref{eqn:dyson_G},\ref{eqn:dyson_H},\ref{eqn:diagramDysonG},\ref{eqn:diagramDysonH})} form a set of closed integral equations that need to be solved self-consistently. This will be computationally demanding, and limitations of numerical methods could obscure the important features we aim to explore. For illustrative purposes, we approximate $\Sigma$ appearing in Eq.~(\ref{eqn:dyson_H}) perturbatively to first order in $\alpha/\Delta_0$. Formally, this is obtained with the following two approximations:

\begin{itemize}
    \item[{(1).}] We evaluate $G$ \emph{perturbatively} to order $\alpha/\Delta_0$:
\end{itemize}
\begin{equation}\label{eqn:final_dyson_G}
    \begin{fmffile}{fmf_choice_of_g}
    \parbox{40pt}{
    \begin{fmfgraph*}(40,40)
        \fmfset{arrow_len}{6}
        \fmfset{arrow_ang}{20}  
        \fmfleft{i}
        \fmfright{f}
        \fmf{heavy}{i,f}
    \end{fmfgraph*}
     }  \,  \approx \,
     \parbox{40pt}{
     \begin{fmfgraph*}(40,40)
        \fmfset{arrow_len}{6}
        \fmfset{arrow_ang}{20}  
        \fmfleft{i}
        \fmfright{f}
        \fmf{fermion}{i,f}
     \end{fmfgraph*}
      } \,+\,  \parbox{120pt}{
      \begin{fmfgraph*}(120,30)
        \fmfset{arrow_len}{6}
        \fmfset{arrow_ang}{20}  
        \fmfleft{i}
        \fmfright{f}
        \fmf{fermion,tension=2}{i,v1}
        \fmf{fermion,tension=2}{v2,f}
        \fmf{fermion}{v1,v2}
        \fmf{plain,foreground=(1,,0,,0),left}{v1,v2}
     \end{fmfgraph*}
      }
    \end{fmffile}
\end{equation}

This approximate form of $G$ allows us to write $\Sigma$ up to order $\alpha/\Delta_0$:
\begin{equation}
    \Sigma[H] = \chi_{11} + \Sigma^{\text{1loop}}[H] + \mathcal{O}((\alpha/\Delta_0)^2) \, , 
\end{equation}
where: 

\begin{eqnarray}
    &&\begin{fmffile}{fmf_chi11}
            \parbox{40pt}{
\begin{fmfgraph*}(40,40)
\fmfleft{l}
\fmfright{r}
\fmfset{arrow_len}{7}
\fmfforce{(0w,0.5h)}{l}
\fmfforce{(1w,0.5h)}{r}
\fmfforce{(0.15w,0.5h)}{v1}
\fmfforce{(0.85w,0.5h)}{v2}
\fmf{plain,foreground=(1,,0,,0)}{l,v1}
\fmf{plain,foreground=(1,,0,,0)}{v2,r}
\fmf{fermion,right=1}{v1,v2,v1}
\end{fmfgraph*}
} 
\end{fmffile} = \chi_{11}(\iq,\iw) \nonumber\\
&=& -\sum_{\iv, \ik}\Tr[G_{0,\ik}(\iv) \sigma_1 G_{0,\ik+\iq}(\iv+\io) \sigma_1]\, . 
\end{eqnarray}

{We define the one-loop self-energy \(\Sigma^{\text{1loop}}\) as those with one integral over the collective mode momentum; it is of order $\alpha/\Delta_0$ as argued in Eq.~(\ref{eqn:MagnitudeOfIntegral}):}
\begin{eqnarray}\label{eqn:self_energy_LW}
    \Sigma^{\text{1loop}}[{H}] \, \, &=& \, \, 
\begin{fmffile}{fmf_sigma_1loop}
  \,\, 
  2\,
\parbox{40pt}{
\begin{fmfgraph*}(40,40)
\fmfleft{l}
\fmfright{r}
\fmfset{arrow_len}{7}
\fmfforce{(0w,0.1h)}{l}
\fmfforce{(1w,0.1h)}{r}
\fmfforce{(0.2w,0.2h)}{v1}
\fmfforce{(0.8w,0.2h)}{v2}
\fmfforce{(0.2w,0.8h)}{v4}
\fmfforce{(0.8w,0.8h)}{v3}
\fmf{plain,foreground=(1,,0,,0)}{l,v1}
\fmf{plain,foreground=(1,,0,,0)}{v2,r}
\fmf{fermion,right=0.42}{v1,v2,v3,v4,v1}
\fmf{plain,foreground=(1,,0,,0)}{v3,v4}
\end{fmfgraph*}
}
+ \,1  \quad
\parbox{40pt}{
\begin{fmfgraph*}(40,40)
\fmfleft{l}
\fmfright{r}
\fmfforce{(-0.1w,0.5h)}{l}
\fmfforce{(1.1w,0.5h)}{r}
\fmfforce{(0.07573w,0.5h)}{v1}
\fmfforce{(0.5w,0.07573h)}{v2}
\fmfforce{(0.92426w,0.5h)}{v3}
\fmfforce{(0.5w,0.92426h)}{v4}
\fmfset{arrow_len}{7}
\fmf{fermion,right=0.42}{v1,v2,v3,v4,v1}
\fmf{plain,foreground=(1,,0,,0)}{l,v1}
\fmf{plain,foreground=(1,,0,,0)}{v3,r}
\fmf{plain,foreground=(1,,0,,0)}{v4,v2}
\end{fmfgraph*}
}
\end{fmffile}
  \quad \nonumber\\
&+&2\,  
\begin{fmffile}{fmf_sigma_1loop_2}
\parbox{80pt}{
 \begin{fmfgraph*}(80,40)
\fmfset{arrow_len}{7}
    \fmfleft{i1,i2,i3}
    \fmfright{o1,o2,o3}
    \fmf{plain,foreground=(1,,0,,0),tension=0.3}{i2,v1}
    \fmf{phantom,tension=0.4}{i1,v2}
    \fmf{phantom,tension=0.4}{i3,v3}
    \fmf{fermion,tension=0.05}{v1,v2,v3,v1}
\fmf{plain,foreground=(1,,0,,0),tension=0.3,left=1/4}{v3,p2}
\fmf{plain,foreground=(1,,0,,0),tension=0.3,left=1/4}{p3,v2}
    \fmf{fermion,tension=0.05}{p1,p2,p3,p1}
    \fmf{phantom,tension=0.4}{p2,o3}
    \fmf{phantom,tension=0.4}{p3,o1}
    \fmf{plain,foreground=(1,,0,,0),tension=0.3}{p1,o2}
  \end{fmfgraph*}
}
\end{fmffile} \, . 
\end{eqnarray}

{The second and third terms are similar to the Maki-Thompson \cite{Maki,Thompson} and Aslamazov-Larkin \cite{Aslamazov_larkin} terms in the study of fluctuation corrections to the conductivity. In our case, instead of the conductivity, we compute the correction of these terms to the collective mode propagator.} It is convenient to define 

$$
H^{-1}_{0,\iq}(\iw) = \frac{2}{V} - \chi_{11}(\iq,\iw) \, , 
$$
which is the RPA level Higgs propagator. This leads to the second approximation:
\begin{itemize}
    \item[{(2).}] We treat $H$ appearing in $\Sigma^{\text{1loop}}$ as $H_0$ which means that we do not solve $\Soneloop$ self-consistently and approximate $\Soneloop[H] \approx \Soneloop[H_0]$
\end{itemize}

The Dyson equation Eq.~(\ref{eqn:dyson_H}) is then rewritten as:
\begin{equation}\label{eqm:final_dyson_higgs}
    H^{-1}_{\iq}(\iw) = H_{0,\iq}^{-1}(\iw) - \Sigma^{\text{1loop}}[H_0]\, ,
\end{equation}
which approximates $\Sigma$ up to order $\alpha/\Delta_0$. 

\subsection{Connection to HS transformation}
While our Dyson equations are obtained from a Luttinger-Ward functional description of superconductivity and can be extended to more sophisticated approximations, the same one-loop results can also be obtained from the conventional HS transformation \cite{Lara_cooperpairVSHiggs,Cea_cdw,Schwarz_2021}. Taking Eq.~(\ref{eqn:yukawa_action2}), expand \(\Delta\) around the vacuum expectation  and integrate out the fermions, the effective action after HS transformation \(S_{HS}\) in the frequency and momentum space is:
\begin{align}
    S_{{HS}} = \int_{q} \frac{(\Delta_{0}+h_{q})^2}{V} + \sum^{\infty}_{n=0}\Tr \frac{(G_{0k} h_{k - k'}\sigma_{1})^n}{n}
\end{align}
where \(q = (\iw,\iq)\) and \(\int_{q} = \int d\omega/2\pi \ d^d\iq/(2\pi)^d\). The trace \(\Tr\) will be taken over the frequency, momentum, and Nambu indices. Truncating the infinite sum up to \(n=4\) gives rise to an effective theory:
\begin{widetext}
    \begin{eqnarray}
    S[h] &=&  \int_{-\infty}^\infty d\tau \frac{2(\Delta_0 h(\tau))}{V} + \sum_{\ik, i\nu} h(\tau)\Tr[G_{0,\ik}(i\nu)\sigma_1] \nonumber \\ 
    &+& \frac{1}{2} \int_q  h(q)\left[ \frac{2}{V} - \chi_{11}(q) \right]h(-q)  - \int_{k,q} \Gamma^{(3)}(k,q) h_{-k} h_{-q} h_{k+q} - \int_{k,q,p} \Gamma^{(4)}(k,q,p)h_{-k}h_{-q}h_{-p}h_{k+q+p} \label{eqn:4th_order_action} \\
    \chi_{ij}(q) &=&  \int_{k'}\Tr[G_{0,\ik+\iq}(\iv + 
\iw)\sigma_i G_{0,\ik}(\iv) \sigma_j] \label{eqn:chi_definition}\\ 
\Gamma^{(3)} &=&  -\frac{1}{3} \int_{k'} \Tr[G_{0,\ik'+\ik}(i\nu + i\omega)\sigma_1 G_{0,\ik'}(i\nu) \sigma_1 G_{0,\ik'-\iq}(i\nu-i\Omega)\sigma_1]\\
\Gamma^{(4)} &= &  -\frac{1}{4} \int_{k'} \Tr[G_{0,\ik'-\ik}(\iv-\io) \sigma_1 G_{0,\ik'+\ik+\iq+\ip}(\iv + \io + i\lambda) \sigma_1 G_{0,\ik'+\iq}(\iv + \io) \sigma_1 G_{0,\ik'}(\iv) \sigma_1]
\end{eqnarray}
\end{widetext}

These define an interacting theory of the scalar field $h$. The same self-energy Eq.~(\ref{eqn:self_energy_LW}) is simply the one-loop Hartree and Fock self-energy of the theory, under the same constraint Eq.~(\ref{eqn:gap_equation_FullGF}) which is equivalent to setting \(\expval{h}=0\).

Further details on the definition of the Higgs propagator at finite momentum and analytic continuation can be found in the supplementary material \cite{SM}. In the next section, we present the numerical solutions to $\Sigma$ and \(H\), and show that indeed \(H(\iw)\) has a pole inside $2\Delta_0$.

\section{Results}\label{sec:results}
In this section, we solve for $\Soneloop$ to one-loop order and examine the effect of QFs on the dressed Higgs propagator Eq.~(\ref{eqn:dyson_H})
\subsection{Modification to the gap equation}
Let us focus on Eq.~(\ref{eqn:final_dyson_G}) first. This equation suggests a modification to the self-consistency equation: 

\begin{equation}\label{eqn:diagrammatic_gap_equation_modified}
\begin{fmffile}{gap_equation}
\frac{2\Delta_0}{V} = \quad \parbox{40pt}{
\begin{fmfgraph*}(40,40)
\fmfset{arrow_len}{7}
    \fmfleft{i}
    \fmfright{j}
    \fmf{fermion,left=1}{i,j,i}
    \fmfdot{i}
    \fmfv{l.a=0,l=$\sigma_1$}{i}
\end{fmfgraph*}} \, + \, 
\parbox{40pt}{
\begin{fmfgraph*}(40,40)
\fmfset{arrow_len}{7}
    \fmfleft{i}
    \fmfright{j}
    \fmftop{t}
    \fmfbottom{{b}}
    \fmf{fermion,left=0.42}{i,t,j,b,i}
    \fmfdot{i}
    \fmf{plain,foreground=(1,,0,,0)}{t,b}
    \fmfv{l.a=0,l=$\sigma_1$}{i}
\end{fmfgraph*}}
\end{fmffile}
\end{equation}  

Including the additional diagram representing QF contributions and substituting \(H_{0\iq}(\iO)\), we find numerically that to one-loop order the gap equation is modified to:

\begin{equation}
    { \frac{2}{V} = \sum_\ik\frac{1}{E_\ik} - c_d\frac{\alpha N_F}{\Delta_0}}\, ,
\end{equation}
where $c_d$ is a numerical factor depending on the spatial dimension \(d\), with $c_2 = 0.3968$ and $c_3 = 0.4106$. The above equation is solved by shifting the superconducting gap:
\begin{equation}\label{eqn:renormalised_gap}
    {\Delta_0 = \Delta_{BCS} - \frac{\alpha c_d}{2}}
\end{equation}
where \(\Delta_{BCS}\) denotes the BCS solution, which solves Eq.~(\ref{eqn:algebraic_gap_equation}). The bare Higgs propagator is modified accordingly:
\begin{eqnarray}
    H^{-1}_{0,\iq=0}(\iw) &=& \frac{2}{V} - \chi_{11}(\iq=0,\iw) \nonumber \\ 
    &=&(4\Delta^2_0 - (\iw)^2)N_F F(\iw) - c_d\frac{\alpha N_F}{\Delta_0}
\end{eqnarray}

In practice, for simplicity, we will not include the \(c_{d}\) dependent term to $H_{0}$ in evaluating $\Soneloop[H_{0}]$. This constant shift to $H^{-1}$ would introduce corrections in $\Soneloop$ that are higher order in the expansion parameter $\alpha/\Delta_0$, and we may safely ignore it for the one-loop results.

\subsection{Numerical solution to $\Soneloop$}

Here we will present the numerically solved $\Soneloop$. Some details will be presented in the supplementary material \cite{SM}, including the definition of finite-$\iq$ propagators and the procedure of analytic continuation we use. We present the final results here, isolating the dimensionless part of the Higgs self-energy, we define:
\begin{equation}
    {\Sigma}^{\text{1loop}}(i\omega/\Delta_0) \coloneqq \frac{\alpha N_F}{\Delta_0} \times \bar{\Sigma}^{\text{1loop}}(i\omega/\Delta_0)\, .
\end{equation}

We are interested in the long-wave limit behavior of the Higgs propagator, whose properties are determined by the $\iq\rightarrow0$ Dyson equation:
\begin{eqnarray}\label{eqn:higgs_inv_dressed}
    &&(N_FH_{\iq=0})^{-1}(\iw) \nonumber \\
    &=& (4\Delta_0^2 - (\iw)^2)F(\iw) 
    - \frac{\alpha\bar\Sigma^{\text{1loop}}(\iw)}{\Delta_0}   - \frac{c_d\alpha }{\Delta_0} \,.
\end{eqnarray}

Although we expand perturbatively in the parameter $\alpha/\Delta_0$, we deem it appropriate to present the results as a function of $\delta = \Delta_0/E_F$ instead, related to $\alpha$ via Eq.~(\ref{eqn:alpha_delta_relation}). This reveals the importance of spatial dimensions on the effects of QFs.

\begin{figure}[h]
    \centering
    \includegraphics[width=1\linewidth]{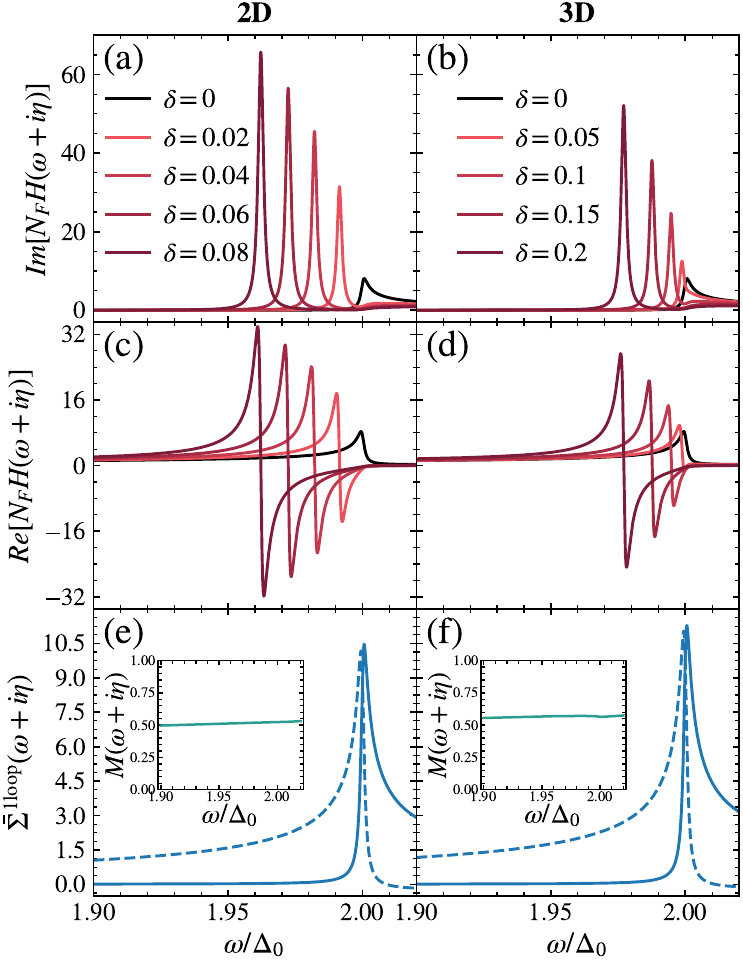}
    \caption{Numerical solutions of Eq.~(\ref{eqn:higgs_inv_dressed}). (a) and (b)---Imaginary parts of the dressed Higgs propagator $H(\omega+i\eta)$ in two and three spatial dimensions, plotted at different values of $\delta$. In two dimensions $\alpha/\Delta_0 \propto \delta$ while in three dimensions $\alpha/\Delta_0 \propto \delta^2$, and hence much higher values of $\delta$ are needed to shift the Higgs mode pole {below} $2\Delta_0$. (c) and (d)---Real parts of the dressed Higgs propagator. (e) and (f)---The real (dashed lines) and imaginary (solid lines) parts of the unitless one-loop self-energy $\bar{\Sigma}^{\text{1loop}}$. In the insets of (e) and (f), we plot the mass correction function $M$ defined in the main text Eq.~(\ref{eqn:mass_correction_function}). We regularize our results with $\eta = 0.001\Delta_0$.}
    \label{fig:modified_higgs}
\end{figure}

Figure (\ref{fig:modified_higgs}) shows the final numerical results of the dressed Higgs propagator and the one-loop self-energy $\Soneloop[H_0]$. The $x$-axis is renormalized with respect to $\Delta_0$ shown in Eq.~(\ref{eqn:renormalised_gap}), which is different from the BCS value $\Delta_{BCS}$. In the subfigures $(a)$, $(b)$, $(c)$, and $(d)$ we present the real and imaginary parts of $H(\omega + i\eta)$. $\delta=0$ represents the RPA level results and is marked in black. In both two and three spatial dimensions, the real part of $H$ never crosses zero for $\omega<2\Delta_0$, which is indicative of a square-root-like divergence $1/\sqrt{4\Delta_0^2 - \omega^2}$. As we tune $\delta$ away from zero, the peak of $H$ slowly shifts to $\omega<2\Delta_0$, and the imaginary parts ($(a)$ and $(b)$) become much sharper. Importantly, we see that for $\delta\neq0$ the real part changes sign at the singularity of $H$, which implies that {the Higgs mode structure is indeed:}
$$
H \sim \frac{1}{\omega_H^2 - (\omega+i\eta)^2}\,
$$ 
near a divergence below the quasiparticle gap {$\omega_H < 2\Delta_0$}. To see how this arises from Eq.~(\ref{eqn:higgs_inv_dressed}), define the \emph{mass correction function}:

\begin{equation}\label{eqn:mass_correction_function}
    M(\io) = \frac{\bar\Sigma^{\text{1loop}}(\iw)}{\Delta_0^2F(i\omega)}\,,
\end{equation}
such that $M(\iw)$ is dimensionless. We plot {the analytically continued} $M(\omega+i\eta)$ and $\Soneloop (\omega + i\eta)$ in Fig. (\ref{fig:modified_higgs}$e$, \ref{fig:modified_higgs}$f$). The imaginary part (solid lines) of $\Soneloop$ is suppressed for $\omega < 2\Delta_0$. This is expected, since $\Sigma$ only contains the generation of Nambu quasiparticle pairs and the creation of $H_0$ propagators, both of which are gapped at $2\Delta_0$. Therefore, inside $\omega<2\Delta_0$ and to first order in $\alpha/\Delta_0$, there are no states for $H$ to decay into, leading to a vanishing imaginary part of $\Soneloop$. Hence $M(\omega + i\eta)$ is real for $\omega<2\Delta_0$, and numerical results suggest that $M(\omega + i\eta)$ is positive and nearly flat, as can be seen in the insets of Fig.~(\ref{fig:modified_higgs}e, \ref{fig:modified_higgs}f). This supports our conjecture Eq.~(\ref{eqn:conjecture}). 

Returning to the Dyson equation Eq.~(\ref{eqn:higgs_inv_dressed}), we can now rewrite it using $M(\iw)$:

\begin{eqnarray}\label{eqn:higgs_dyson_2}
    &&(N_FH_{\iq=0})^{-1} \nonumber \\
    &=& \{4\Delta_0^2- \alpha \Delta_0M(\iw) -  c_d \frac{\alpha \Delta_0}{F(\iw)} -(\iw)^2\} F(\iw)\,.
\end{eqnarray}

Since we know  $M(\omega + i\eta)$ is real and positive-definite for $\omega<2\Delta_0$, and:
$$
1/F(\iw) = \frac{\iw \sqrt{4\Delta_0^2 - (\iw)^2}}{2\arcsin(\iw/2\Delta_0)}
$$
is also regular, real, and positive definite for $\omega < 2\Delta_0$ after analytic continuation. The pole of the dressed Higgs propagator is obtained by solving:
\begin{equation}\label{eqn:equation_for_omega_h}
    (\omega)^2 = 4\Delta_0^2- \alpha \Delta_0M(\omega) -  c_d \frac{\alpha \Delta_0}{F(\omega)}
\end{equation}
whose solution $\omega = \omega_H < 2\Delta_0$ lies inside the pair breaking gap, and vanishes like a Lorentzian {pole}.

For illustrative purposes, it is helpful to provide the $\delta\rightarrow0$ limit solution to the above equation. Since $1/F(2\Delta_0)$ vanishes, the contribution from the $c_d$ term is small in the limit $\omega \rightarrow 2\Delta_0$. Taking $M(\omega) \approx M(2\Delta_0)$ we obtain:
\begin{equation}\label{eqn:simplifed_solution_omega_h_alpha}
    \frac{\omega_H}{\Delta_0} \approx 2 - 0.14\alpha/\Delta_0
\end{equation}
or, written in $\delta$ assuming a parabolic band:
\begin{equation}\label{eqn:simplifed_solution_omega_h}
    \frac{\omega_H}{\Delta_0} \approx \begin{cases}
        2 - 0.39 \delta & \text{two dimensions} \\ 
        2- 0.43\delta^2 & \text{three dimensions}
    \end{cases}
\end{equation}
are the solutions to the Higgs mode pole in the weak coupling limit $\delta\rightarrow 0$. Hence, we conclude that the Higgs mode always appears close, but below the pair breaking gap in $s$-wave superconductors. 

This conclusion is in line with previous studies \cite{park_holstein_BCSBEC,lorenzanaLongLivedHiggsModes2024a,lorenzo_two_particle_self_consistent,cabreraHybridizationAmplitudeMode2025} where the Higgs mode appears significantly below the pair breaking gap in the crossover regime between the Bardeen–Cooper–Schrieffer (weak coupling) and the Bose-Einstein-condensation (strong coupling) limits. We have shown that even in the BCS limit ($\delta \rightarrow 0$), the Higgs mode appears below the gap with distance $\sim \Delta_0\delta^{d-1}$. This significantly enhances the Higgs mode spectral strength.



\section{Response functions}\label{sec:response_functions}
To grasp the influence of QFs on physical observables, we compute two fingerprints where the Higgs mode can appear—the Third Harmonic Generation (THG) \cite{shimano_2020} and the $A_{1g}$ Raman response \cite{devereaux}. {These spectra} are determined by the corresponding nonlinear {susceptibilities} \cite{Lara_cooperpairVSHiggs,Schwarz_2021,rafael_impurity,devereaux}:

\begin{subequations}\label{eqn:nonlinear_susceptibilities}
    \begin{eqnarray}
    \chi_{ij}(\io) &=& - \sum_{\ik.\iv}\Tr[\sigma_i G_\ik(\iv+\io)\sigma_j G_\ik(\iv)]\, ,   \\
     \chi_{PB} &=& \chi_{\gamma\gamma} -\frac{\chi_{\gamma 3}^2}{\chi_{33}}\, ,  \\
     \chi_{H} &=& -\frac{(\chi_{\gamma 1} - \chi_{\gamma 3}\chi_{31}/\chi_{33})^2}{H^{-1} + \chi_{13}^2/\chi_{33}} \label{eqn:Higgs_thg_raman}\, . 
\end{eqnarray}
\end{subequations}

{To include the effect of QFs investigated in the previous section, we use the dressed Higgs propagator $H$ computed from Eq.~(\ref{eqn:higgs_inv_dressed})}. {Here}, $\chi_{PB}$ and $\chi_H$ are the Coulomb-screened susceptibility contributions from quasiparticle pair-breaking (PB) and the Higgs mode, respectively. $\chi_{\gamma j}$ means that one of the vertices $\sigma_i$ is replaced by the light-matter interaction vertex $\gamma \sigma_3$ {(likewise $\chi_{\gamma \gamma}$ replaces both vertices)}. Here, the vertex $\gamma$ can be either the THG or the Raman vertex, defined as 
\begin{subequations}\label{eqn:thg_and_raman_gamma}
    \begin{eqnarray}
        \gamma_{THG}(\theta_P) &=& (\cos(\theta_P)^2\partial_{k_x}^2 +\sin(\theta_P)^2\partial_{k_y}^2 \nonumber\\
        &+&\sin(2\theta_{P})\partial_{k_{x}}\partial_{k_{y}}) \epsilon_{\ik}\, , \\ 
        \gamma_{A_{1g}} &=& \frac{1}{2}(\partial_{k_x}^2+\partial_{k_y}^2)\epsilon_\ik \, ,
    \end{eqnarray}
\end{subequations}
with $\theta_P$ being the polarization angle of the incoming light pulse relative to the crystal $x$-axis. We use a tight-binding model with nearest-neighbor {and} {next-to-nearest-neighbor} hoppings ${\xi_\ik} = -2t (\cos k_x + \cos k_y)-4t'(\cos k_x  \cos k_y) - \mu$ with $\mu = 1t$, $t'=0.35t$. We compute the susceptibilities using the Cubature adaptive integration algorithm \cite{genzRemarksAlgorithm0061980,berntsenAlgorithm698DCUHRE1991}. The results are summarized in Fig.~(\ref{fig:Raman_THG}). In the panels (c), (d), and (e), we present the Raman $A_{1g}$ and THG susceptibilities. These were obtained from Eq.~(\ref{eqn:nonlinear_susceptibilities}, \ref{eqn:thg_and_raman_gamma}) after analytic continuation $i\omega\rightarrow \omega+i\eta$ $\chi(\iw) \rightarrow \chi(\omega+i\eta) = \chi^R(\omega)$. The Raman susceptibility is proportional to the imaginary part of the response function $\chi_{A_{1g}}$, while the THG susceptibility will be proportional to $\abs{\chi^R(\omega)}$.

Before we discuss QFs, we briefly explain the RPA level results. The Higgs mode couples quadratically to light and is therefore Raman-active. However, unlike the THG response, where the relative intensities of PB and Higgs contributions were analysed extensively \cite{Lara_cooperpairVSHiggs,Schwarz_2021,rafael_impurity,tsuji_impurity,shimano_2020}, similar comparisons with Raman responses are much rarer. The key observation is that if \(t'=0\), the $A_{1g}$ Raman vertex is proportional to the THG vertex evaluated at $\theta_P = \pi/4$, both are isotropic and proportional to \(\gamma_{THG/A_{1g}} \propto \xi_{\ik}\). {In this case, screening from charge conservation will strongly suppress this isotropic component \cite{maitiConservationLawsVertex2017} and remove the divergence from the PB response \cite{Lara_cooperpairVSHiggs} in both the THG and Raman responses. The case is different for the Higgs mode, since \(\gamma_{THG/A_{1g}} \propto \xi_{\ik}\) allows finite coupling between the Higgs mode and light, and the Higgs propagator \(H\) remains diverging, there will be a Higgs-dominant THG response for \(t'=0\) at \(\theta_{P}=\pi/4\). This was pointed out in reference \cite{Lara_cooperpairVSHiggs}. Consequently, the same mechanism is expected to ensure a Higgs-dominant $A_{1g}$ Raman response. 


If we include the next nearest neighbour hopping \(t'\), the THG vertex \(\gamma_{THG}\) will be a mixture of \(A_{1g}+B_{1g} + B_{2g}\) Raman vertices:
\begin{subequations}
    \begin{align}
\gamma_{THG}  & = \gamma_{A_{1g}} + \cos(2\theta_{P}) \gamma_{B_{1g}} + \sin(2\theta_{P}) \gamma_{B_{2g}}         \\
\gamma_{B_{1g}}  & = \frac{1}{2}(\partial_{k_{x}}^{2}-\partial_{k_{y}}^{2})\epsilon_{\ik} \\
\gamma_{B_{2g}} & =(\partial_{k_{x}}\partial_{k_{y}}) \epsilon_{\ik} \,. 
    \end{align}
\end{subequations}
The PB THG susceptibility \(\chi_{PB}^{THG}\) thus decomposes into different Raman symmetries, but the isotropic Higgs mode only contains the \(A_{1g}\) component. In other words:

\begin{subequations}
\begin{align}
    \chi_{PB}^{THG}  & = \chi_{PB}^{A_{1g}} + \cos(2\theta_{P})^{2}\chi_{PB}^{B_{1g}} + \sin(2\theta_{P})^{2}\chi_{PB}^{B_{2g}}  \\
    \chi^{THG}_{H}  & = \chi^{A_{1g}}_{H} \, . 
\end{align}    
\end{subequations}

The dominance of the pair-breaking THG response over the Higgs mode contribution in a clean system \cite{Lara_cooperpairVSHiggs,Schwarz_2021,tsuji_impurity} is therefore a consequence of the \(B_{1g}+B_{2g}\) responses dominating over the screened \(A_{1g}\) response. In Raman measurements, where one can isolate the \(A_{1g}\) contributions, screening reduces the pair-breaking response, leading to a Higgs-dominant response as explained above for \(t'=0\). Including \(t'\) provides an anisotropic component to the \(A_{1g}\) Raman vertex and recovers the divergence of the pair-breaking response. The intensity ratio of PB and Higgs responses then depends on the details of the model parameters. Fig.~(\ref{fig:Raman_THG}) presents the results on the tight-binding band. {We see in Fig.~(\ref{fig:Raman_THG}c,\ref{fig:Raman_THG}d), while the Higgs mode is still subdominant without QFs, it constitutes a significant portion of the peak at $2\Delta_0$ in the Raman response.



\begin{figure}
    \centering
    \includegraphics[width=\linewidth]{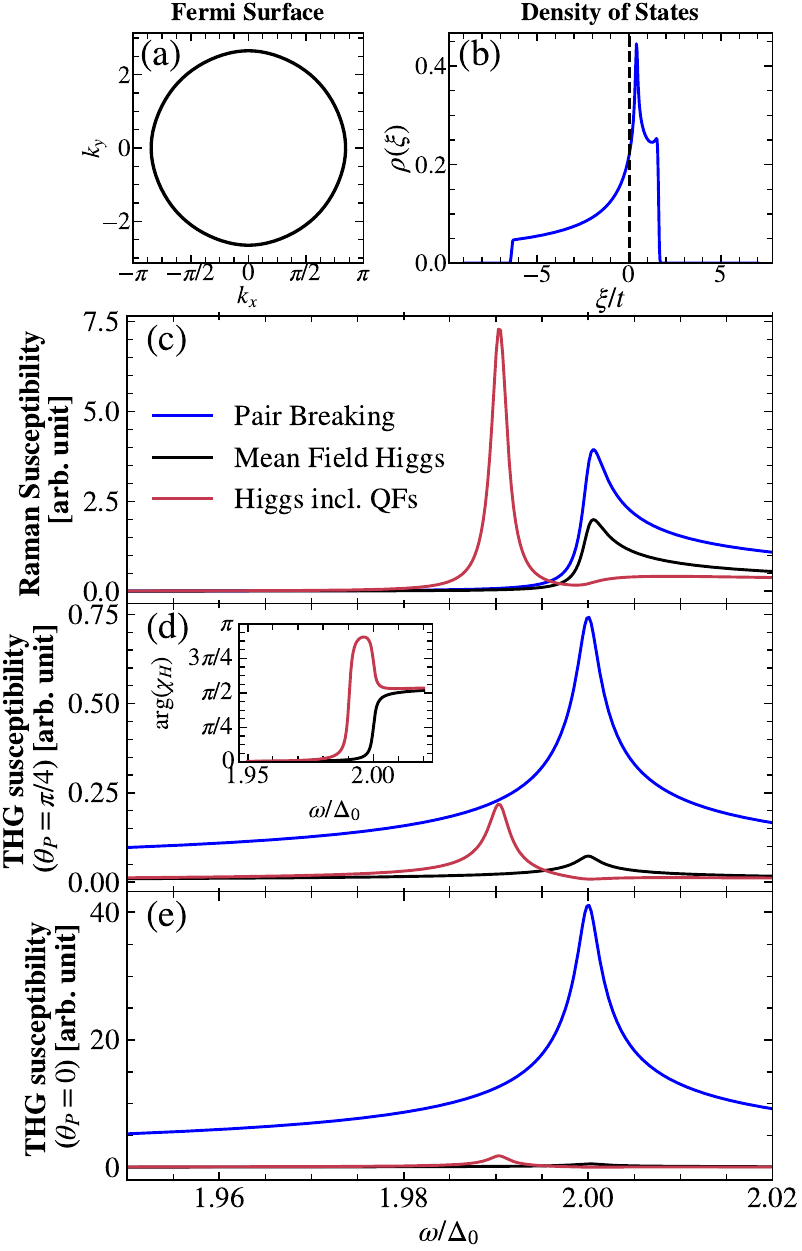}
    \caption{ Panels (a) and (b)---The Fermi surface and the density of states of our model. We have marked the Fermi level $\xi=0$ with a dashed line in (b). Panels (c)---$A_{1g}$ Raman susceptibility $\Im{\chi^R}$. (d)---THG susceptibility $\abs{\chi^R}(\theta_P=\pi/4)$. Inset of (c)---The THG phase jump, there is clearly a region between $\omega_H$ and $2\Delta_0$ where the phase jumps to $\pi$ instead of the usual $\pi/2$ (e)---THG susceptibility $\abs{\chi^R}(\theta_P=0)$. For this model $U/t = 1.35$, $\Delta_0/t = 0.15$. This gives an effective {$\alpha/\Delta_0 = 0.07$}. Plots are regularized with $\eta/\Delta_0 = 0.001$}
    \label{fig:Raman_THG}
\end{figure}

We now introduce fluctuation effects beyond RPA. Corrections from QFs should not influence QP responses qualitatively. The square root divergence of QP response is due to a continuum of particle-like excitations, and we do not expect our weak QFs to ruin this quasiparticle picture. On the other hand, QFs change the Higgs propagator qualitatively. The most important correction from QFs can thus be captured by replacing the Higgs propagator appearing in Eq.~(\ref{eqn:nonlinear_susceptibilities}) with the dressed Higgs propagator Eq.~(\ref{eqn:higgs_inv_dressed}). {We estimate the value of $\alpha$ using the parameters of our model: {$E_F = -\xi_{\ik=0} = 6.4t$, $\delta \approx 0.023 $, $\alpha/\Delta_0 \approx 0.07$}. The red lines in Fig.~(\ref{fig:Raman_THG}c,\ref{fig:Raman_THG}d,\ref{fig:Raman_THG}e) shows the corresponding results. Importantly, the pole of the Higgs propagator appears at roughly $\omega_H = 1.99\Delta_0$ with a much higher spectral weight. }

\emph{Raman}---.For the Raman spectrum, the situation is similar to that found in 2H-NbSe$_2$ \cite{sooryakumarRamanScatteringSuperconductingGap1980,meassonAmplitudeHiggsMode2014}. {This material is notable for the coexistence of superconductivity and charge-density-wave (CDW) order. The observability of the Higgs mode is partially attributed to the distinction between $\omega_H$ and the quasiparticle gap in this material \cite{Cea_cdw,Schwarz_2021}. In the language of this work, in NbSe$_2$, the $F(\omega)$ function peaks at {a} frequency above $2\Delta$ and the Higgs mode reveals its pole. This is the same mechanism that allows QFs to enhance our spectral signal---revealing the Higgs mode pole. In other words, QFs could play the role of CDW, leading to an experimentally observable $s$-wave Higgs pole. This is clearly shown in  Fig.~(\ref{fig:Raman_THG}c) where a sharp peak appears slightly below the gap. Consequently, we expect discrepancies between {the} Raman A$_{1g}$ peak position and the pair excitation gap $2\Delta_0$. The latter can be measured directly via the scanning tunneling microscope (STM). 

\emph{THG}---.In the THG spectrum, we have found again a stronger THG resonance peak for the Higgs mode. This would be clearly experimentally observable for the $\theta_P = \pi/4$ orientation, as shown in Fig.~(\ref{fig:Raman_THG}). For $\theta_P=0$ polarization, as expected from clean-limit calculations, both THG signals from the Higgs modes (red and black lines) are subdominant compared to QP excitations (blue lines). {Nonetheless, we have shown that the Higgs mode propagator has a significantly stronger singularity by including QFs. It is well-known that impurity enhances the optical response of the Higgs mode \cite{murotani_impurity,silaev_impurity,tsuji_impurity,seibold_impurity,rafael_impurity}; combining with our result, we believe that the Higgs mode can dominate the THG response at a lower impurity level than in the literature. A qualitative change is also expected to appear in the phase of the THG signal, as illustrated by the inset in Fig.~(\ref{fig:Raman_THG}d). For a square-root-like divergence, a phase jump of $\pi/2$ is expected as frequency goes from below to above the divergence \cite{Schwarz_2021}. With QFs, the Higgs mode gains a pole structure rather than a square-root divergence, implying that between $\omega_H$ and $2\Delta_0$ the THG phase of the Higgs mode will jump to $\pi$ rather than $\pi/2$. This can be measured by a THG driving sweep near $2\Delta_0$, and search for any region where the THG phase jumps above $\pi/2$. Another possible test, which was not computed theoretically in this work, is a high-resolution temperature-resolved phase measurement of THG signal. The deviation between $\omega_H$ and $2\Delta_0$ is a consequence of QFs and is expected to remain even at finite temperature. We may also measure similar phase jump if one tunes the temperature $T$ such that the driving frequency $\Omega$ of THG satisfies $2\Omega \lesssim 2\Delta_0(T)$; a phase jump larger than $2\pi$ would be expected.

\begin{table*}[t!]
\centering
\caption{List of some superconductors and their corresponding $\alpha$\label{table1}. $\Delta_0$, size of the single-particle spectrum gap; $N_F$ density of states at the Fermi surface; $\xi_{GL}$, Ginzburg-Landau coherence length; $\xi_0$, coherence length defined in this work; $\alpha$, parameter that measures the strength of QFs, defined in Eq.~(\ref{eqn:alpha_definition}); $\alpha/\Delta_0$, this ratio can be used to infer the position of $\omega_H$ using Eq.~(\ref{eqn:simplifed_solution_omega_h_alpha}). }
\vspace{2mm} 

\begin{tabular*}{\textwidth}{@{\extracolsep{\fill}} c c c c c c c c @{}}
\hline\hline 
3D Superconductors & $\Delta_0$ & $N_F$ & $\xi_{GL}$ & $\xi_0$ & $\alpha$& $\alpha/\Delta_0$& Ref \\
 & (meV) & (meV$^{-1}$nm$^{-3}$spin$^{-1}$) & (nm) & (nm) & (meV)& &  \\
\hline
NbN   & 2.2  & $0.825-1.19\times10^{-2}$  & $4.30-5.02$  & $3.90-4.56$   & 0.024  & 0.01& \cite{chockalingamSuperconductingPropertiesHall2008,komenouEnergyGapMeasurement1968}    \\
MgB$_2$   & $2.2$($\pi$) \ 5.5($\sigma$)   & $1.24 \times10^{-2}$ & $4.9$  & $4.44$  & $0.016$   &0.008($\pi$) \ 0.004($\sigma$)& \cite{wangSpecificHeatSuperconducting2001,tsudaDefinitiveExperimentalEvidence2003}     \\
Nb$_3$Sn  & $1.98$   & $5.4\times10^{-2}$   & $11.5$   & $10.4$   & $3\times 10^{-4}$   & $1.5\times10^{-4}$& \cite{wangSpecificHeatSuperconducting2001}    \\
FeSe &$0.25$(s)  & $5.03 \times 10^{-2}$ & $5$(ab)   & $4.53$(ab)  & 0.012& 0.0007(s)  & \\ 
&  1.67(es)\footnote{Here s stands for $s$-wave, and es stands for extended $s$-wave.}&&$1.5$(c)\footnote{ab stands for $ab$-plane, and c for $c$-plane.} &  $1.36$(c) &&0.05(es)&\cite{jiaoSuperconductingGapStructure2017,shibauchiExoticSuperconductingStates2020}\\ 

\hline \hline
2D Superconductors & $\Delta_0$ & $N_F$ & $\xi_{GL}$ & $\xi_0$ & $\alpha$ & $\alpha/\Delta_0$& Ref \\
 & (meV) & (meV$^{-1}$nm$^{-2}$spin$^{-1}$) & (nm) & (nm) & (meV) & &  \\
 \hline
 Ultrathin NbN($d=3.2$nm\footnote{This includes both the thin NbN and the oxidized Nb$_2$O$_5$ layer})& 1.55\footnote{Estimated from penetration depth}& 0.0374\footnote{Obtained by multiplying 3D $N_F$ with the  thickness}& $\sim 5$\footnote{Not given explicitly in \cite{semenovOpticalTransportProperties2009,kamlapureMeasurementMagneticPenetration2010}, only gave a range between $4$nm to $6$nm depending on the samples}&$\sim4.5$& 0.104& 0.07& \cite{semenovOpticalTransportProperties2009,kamlapureMeasurementMagneticPenetration2010}\\ 
Monolayer FeSe/SrTiO$_3$ & 9 & $0.5 \times 10^{-2}$ & $2.45$& $2.22$& 3.22 & 0.36&\cite{fanPlainSwaveSuperconductivity2015,huangSuperconductingFeSeMonolayer2021}\\
MAtBG &0.9& $>1.2\times 10^{-4}$ \footnote{This lower bound is estimated by taking the superlattice density $n_s$ and divide it  typical flat-band width $\sim 10$meV }&52nm & 47.2& $0.028<$&$0.03<$& \cite{caoUnconventionalSuperconductivityMagicangle2018a,ohEvidenceUnconventionalSuperconductivity2021a}\\
\hline
\end{tabular*}
\end{table*}

\section{{Candidate materials for experimental verification}}

In this section, we {comment on the material candidates for experimental validation.}

    


We {first} point out the difference between ``strong-coupling superconductors'' and the notion of QFs as defined in this work. This is generally measured relative to the BCS universal relation between the zero temperature gap $\Delta_0$ and the critical temperature $T_c$: $2\Delta_0/k_BT_c = 3.53$. In this sense, NbN with $2\Delta_0/k_B T_c \sim 4-4.4$ \cite{komenouEnergyGapMeasurement1968,chockalingamTunnelingStudiesHomogeneously2009} is {medium- to strong-coupling}. While this is indeed a notion of strong correlation in quantum materials, it \emph{does not} translate to strong QFs in our work. One can see from Table.~(\ref{table1}) that $\alpha/\Delta_0\sim 0.01$ for NbN, leading to vanishingly small QF contribution. Another counter-example is the magic-angle twisted bilayer graphene (MAtBG) where, despite the large tunneling gap $2\Delta_0(T=0)/k_BT_c \sim 25 $ \cite{ohEvidenceUnconventionalSuperconductivity2021a}, we do not expect strong QF corrections ($\alpha/\Delta_0 \sim 0.03$) as shown in Table.~(\ref{table1}). This is because of the quantum geometric contribution to the coherence length in MAtBG \cite{huAnomalousCoherenceLength2025a,ohRoleQuantumGeometry2025}, leading to an anomalously large $\xi_{GL} \sim 52$nm, suppressing the QF contribution. Estimating fluctuation contribution thus requires the knowledge of $\alpha/\Delta_0$ and cannot be inferred simply from $2\Delta_0/k_BT_c$.

The spatial dimension also plays an important role for QFs, as demonstrated in the previous sections. Due to the relation $2-\omega_H/\Delta_0 \propto \delta^{d-1}$ we expect QFs to be significantly more prominent in 2D compared with 3D superconductors. Indeed, Table.~(\ref{table1}) shows that typical examples of strongly-correlated superconductors NbN and Nb$_3$Sn have small $\alpha/\Delta_0$. This persists even if we search beyond $s$-wave symmetry. MgB$_2$ and FeSe \footnote{For FeSe, the coherence lengths are different in the ab-plane and c-plane. Hence we estimate $\alpha = K_d/(N_F\xi_{ab}^2\xi_c)$. } are both strongly-correlated two-band superconductors, yet they also have $\alpha/\Delta_0$ too small to be observed experimentally. In 2D the situation is more optimistic. Ultrathin film of NbN \cite{semenovOpticalTransportProperties2009,kamlapureMeasurementMagneticPenetration2010} with thickness $d\approx 3.2$nm shows $\alpha/\Delta_0=0.07$. This translates to roughly $2\Delta_0 - \omega_H \approx 0.01$meV, hard to resolve experimentally. But for monolayer FeSe on SrTiO$_3$ we have $\omega_H - 2\Delta_0 \approx 0.54$meV. This value is slightly larger than that predicted by Eq.~(\ref{eqn:simplifed_solution_omega_h_alpha}) due to the contribution from the last term in Eq.~(\ref{eqn:equation_for_omega_h}). In the language of Raman shift, we will see a Lorenzian-like peak at $0.43$cm$^{-1}$ below the gap probed with STM. We therefore propose that monolayer FeSe on SrTiO$_3$ is an appropriate candidate to explore the in-gap Higgs mode. 

Very generally, it will be interesting to compare the ``single particle'' gap of a superconductor with the gap in the response functions. Experimental techniques such as STM or angle-resolved photoemission spectroscopy can measure the spectral gap, the gap of exciting a single quasiparticle, whose information is encoded in the Green's function $G_\ik$. On the other hand, response functions such as the Raman susceptibility is the combination of multiple $G_\ik$, with contributions from the collective mode. Our work suggested that the Higgs mode appears below the quasiparticle spectral gap due to QFs. Consequently, this indicates a deviation between gaps measured by, e.g., STM and Raman scattering may appear different due to flucutation corrections. Similar comparisons between STM and optical conductivity gaps of NbN across different levels of disorders were made in \cite{shermanHiggsModeDisordered2015a} and found different gaps. {Since $\xi_0$, and thus $\alpha$, is impurity dependent \cite{chockalingamSuperconductingPropertiesHall2008}, we propose similar measurements with STM and Raman spectroscopy at different disorder, where a Higgs mode should emerge in the Raman spectrum, below the gap measured by STM.

\color{black}

\section{Conclusions and outlooks}\label{sec:conclusion}

We investigated the impact of QFs on the theory of superconductivity and the Higgs mode. While the bare Higgs propagator exhibits a square-root divergence at $\omega_H = 2\Delta_0$ due to its overlap with QP excitation continuum, QFs shift the Higgs mode pole downward, resulting in a well-defined, long-lived mode.}
{The parameter $\alpha$ appears due to the natural ultraviolet cutoff in the collective mode momentum. Physically, the Higgs mode is only well-defined for distances larger than the average Cooper pair size, akin to how the electron momentum is bounded by the details of the lattice. We claim that
$$
\alpha = \frac{K_d}{N_F\xi_0^d}
$$
measures the size of QFs, and is connected to experimentally accessible quantities $N_F$ and $\xi_0$. Our theory, therefore, connects the influence of QFs with observable physical quantities, offering a pathway for experimental validation. This is more prominent in the landscape of 2D superconductors, where the impact of QFs can be experimentally accessible thorugh either a high-resolution temperature-resolved phase measurement of THG signal in superconductors, or by comparing the Raman and STM spectra. Since $\xi_0$, and hence $\alpha$, depend on impurities \cite{shermanHiggsModeDisordered2015a}, we believe comparing the same sample at different impurity levels should also be a venue toward experimental detection of an in-gap Higgs mode.  }

One possible extension of this work is a systematic study of the superconducting Higgs mode in the context of Raman scattering, including different gap symmetries. In $d$-wave superconductors such as cuprates, the interactions are much stronger than those found in conventional ones. QFs should therefore play a more vital role in those systems, and could significantly modify the Raman spectrum.

{Going beyond $s$-wave superconductors, our results may shed light on other symmetry-breaking states such as charge density wave, quantum (anti-)ferromagnets, or cold atom condensates. Moreover, we argue that  QFs are essential in describing cuprate superconductors, where both $N_F$ and $\xi$ are small, leading to large $\alpha$ and the consequent breakdown of mean-field description. }

It is also interesting to extend the techniques developed in this work to finite temperature. For example, near the critical temperature \(T_{c}\) fluctuations on top of the mean-field can be crucial in describing the nature of the phase transition.

\section{Acknoledgement}
We would like to thank Walter Metzner for the critical reading of this work and for the fruitful discussions. We would also like to thank Lukas Debbeler, Steffen Bollmann, Jakob Dolgner, Paulo Forni, Henrik M\"{u}ller-Groeling, Silvia Neri, Robin Scholle, and Yannis Ulrich for useful discussions. We thank the Max Planck-UBC-UTokyo Center for Quantum Materials for fruitful collaborations and financial support. N.T. acknowledges support by JST FOREST (Grant No. JPMJFR2131) and JSPS KAKENHI (Grant Nos. JP24H00191, JP25H01246, JP25H01251).

\bibliography{ref}


\end{document}



\title{Supplemental Material}



  \author{S. Tian}
  \affiliation{Max Planck Institute for Solid State Research, Heisenbergstraße 1, D-70569 Stuttgart, Germany}

  \author{N. Tsuji}
  \affiliation{Department of Physics, The University of Tokyo, Hongo, Tokyo 113-8656, Japan}
  \affiliation{RIKEN Center for Emergent Matter Science (CEMS), Wako 351-0198, Japan}
  \affiliation{Trans-Scale Quantum Science Institute, The University of Tokyo, Hongo, Tokyo 113-8656, Japan}

  \author{D. Manske}
  \affiliation{Max Planck Institute for Solid State Research, Heisenbergstraße 1, D-70569 Stuttgart, Germany}


\date{\today}



\maketitle



In this supplemental material, we present some of the derivation of the effective action for the Higgs mode in superconductors, ignoring the dynamics of the phase mode.
\section{Hubbard Stratonovich Transformation}
The starting point of discussion will be the Hubbard model on a lattice:

\begin{eqnarray}
    \mathcal{H} &=& \sum_{ij \sigma} (t_{ij} c^\dagger_{i \sigma} c_{j \sigma}-\mu\delta_{ij}c^\dagger_{i\sigma}c_{j\sigma}) -V\sum_{i} c^\dagger_{i\uparrow} c^\dagger_{i \downarrow} c_{i \downarrow} c_{i \uparrow} \, ,  
\end{eqnarray}
where the structure of the underlying lattice can be arbitrary. $V>0$ is the point-range attraction, $c_i$ $(c_i^\dagger)$ is the fermionic creation (annihilation) operator at site $i$, with hopping amplitude $t_{ij}$. $\mu$ is the chemical potential. The partition function $\mathcal{Z}$ of this Hamiltonian is:

\begin{equation}
\begin{aligned}
    \mathcal{Z} &= \int \mathcal{D}[c,c^\dagger] e^{-S[c,c^\dagger]} \,, \\
    S &= \int \mathcal{L} d\tau = \int d\tau \sum_{i \sigma} c^\dagger _{i \sigma} \partial_\tau c_{i \sigma} + \mathcal{H}\, .
\end{aligned}
\end{equation}

The four-point interaction can be dealt with by the Hubbard-Stratonovich transformation \cite{benfattoLowenergyPhaseonlyAction2004,ceaNonlinearOpticalEffects2016,haenelTimeresolvedOpticalConductivity2021}:
\begin{eqnarray}
    &&\exp{V\sum_{i} c^\dagger_{i\uparrow} c^\dagger_{i \downarrow} c_{i \downarrow} c_{i \uparrow}} \nonumber\\
    &=& \int \mathcal{D}[\Delta,\Delta^\dagger] \exp{\sum_i\frac{\Delta_i \Delta_i^\dagger}{V} + \Delta_i c^\dagger_{i\uparrow}c^\dagger_{i\downarrow} + \Delta_i^\dagger c_{i\downarrow} c_{i \uparrow}}\,.
\end{eqnarray}
 The action thus becomes quadratic in the fermionic operators $c$ $(c^\dagger)$ at the expense of introducing a complex scalar field $\Delta$:

\begin{eqnarray}\label{eqn:action_0}
    S[c,c^\dagger,\Delta,\Delta^\dagger] &=& \int d\tau \sum_{ij\sigma} c^\dagger_i(\tau)(\partial_\tau\delta_{ij} +t_{ij}-\mu \delta_{ij})c_{j\sigma}(\tau)   \nonumber\\ &&+\sum_i\frac{\Delta_i^\dagger(\tau)\Delta_i(\tau)}{V} + \Delta_i(\tau) c^\dagger_{i\uparrow}(\tau)c^\dagger_{i\downarrow}(\tau)+\Delta^\dagger_i(\tau) c_{i\downarrow}(\tau)c_{i\uparrow}(\tau)
\end{eqnarray}

Notice that this theory has a broken symmetry phase, where $\Delta$ acquires a non-zero vacuum expectation value. Use the notation:
\begin{equation}
\expval{\dots} = \frac{\int\mathcal{D}[c,c^\dagger,\Delta,\Delta^\dagger] (\dots) e^{-S}}{\int\mathcal{D}[c,c^\dagger,\Delta,\Delta^\dagger]e^{-S}}, , 
\end{equation}
we can investigate the extremum of this action:
\begin{equation}
    \frac{\delta \ln\mathcal{Z}}{\delta \Delta_i^\dagger} =0\Rightarrow \expval{c_{i\downarrow}c_{i\uparrow}} =- \expval{\Delta_i}/V \, ,
\end{equation}
We are interested in a homogeneous solution where $\expval{\Delta_i} = \Delta_0$ is a constant, whose complex phase can be chosen to be zero without loss of generality, but such is not true for $\Delta_i$, the complex field. The path integral still integrates over all possible field configurations of $\Delta_i$, including complex ones: $\Delta_i = \abs{\Delta_i}e^{i\theta_i}$. 

\section{Goldstone mode in charged systems}

In a condensed matter system, the phase mode combines with the longitudinal part of the gauge field and becomes the phase-plasmon mode with a pole at the plasma frequency. For example, as shown in \cite{benfattoLowenergyPhaseonlyAction2004} the action of the phase mode in the presence of Coulomb interaction becomes:
\begin{equation}\label{eqn:phase_action}
    S[\theta] = \frac{1}{8} \sum_{i\Omega,\iq} \iq^2\left[\frac{1}{4\pi e^2}\Omega^2 + D_0 \right]\theta(\io,\iq)\theta(-\iw,-\iq)\,,
\end{equation}
here $D_0=\rho/m$ is the phase stiffness with $\rho$ being the particle density and $m$ being the effective mass of electrons. In three dimensions, for example, the phase mode has a pole at the plasma frequency $\omega_p^2 = \frac{4\pi e^2\rho}{m}$, on the order of eVs. In two dimensions, the phase-plason mode is not gapped, but the dynamics remain at a much higher energy scale. Hence, the dynamics of the phase is not expected to contribute to computing loop corrections on the energy scale of $\Delta_0$, even though the phase-plasma mode is still vital in preserving charge conservation. We will therefore ignore the phase-plasmon mode in calculating self-energy corrections to the Higgs mode.

\section{Higgs mode at finite momentum}
Consider the Higgs propagator at the RPA level theory:
\begin{equation}
    H_{0,\iq}(\iw) = \frac{2}{V} - \chi_{11}(\iq,\iw)
\end{equation}
utilizing the BCS gap equation $2/V  = \sum_\ik1/E_\ik$, this equation can be converted into:
\begin{eqnarray}
  H_{0,\iq}(\iw) &=& \frac{2}{V} - \sum_\ik  \frac{E_-+E_+}{E_-E_+}\frac{-\Delta_0^2 + E_-E_+ + \xi_-\xi_+}{(E_-+E_+)^2 - (\io)^2} \nonumber \\ 
  &=& \frac{1}{2}\sum_\ik \frac{E_-+E_+}{E_-E_+} \frac{4\Delta_0^2 - (\xi_--\xi_+)^2 - (\io)^2}{(E_-+E_+)^2 - (\io)^2}
\end{eqnarray}
where $\xi_{\pm} = \xi_{\ik \pm \iq/2}$ and $E_\pm = \sqrt{\xi_\pm^2 + \Delta_0^2}$. Assuming that $\abs{\iq}\ll k_F$, we expand $$\xi_{\ik \pm \iq/2} \approx \xi_\ik \pm  \frac{\ive{v}_F \cdot\iq}{2} = \xi_\ik \pm \frac{v_F q \cos(\theta)}{2}\, .$$
where $v_F = \partial_{\ik}\xi_\ik|_{k = k_F}$ is the Fermi velocity. The Higgs propagator at finite momentum is therefore:
\begin{eqnarray}
    H_{0,\iq}(\io) &=& \frac{1}{2} N_F\int_\infty^\infty d\xi \int_0^{2\pi} \frac{d \theta}{2\pi}\frac{E_-+E_+}{E_-E_+} \frac{4\Delta_0^2 - (\xi_--\xi_+)^2 - (\io)^2}{(E_-+E_+)^2 - (\io)^2} \quad (\text{2D systems}) \label{eqn:higgs_2D} \\ 
    H_{0,\iq}(\io) &=& \frac{1}{2} N_F\int_\infty^\infty d\xi \int_{-1}^{1} \frac{d\cos(\theta)}{2}\frac{E_-+E_+}{E_-E_+} \frac{4\Delta_0^2 - (\xi_--\xi_+)^2 - (\io)^2}{(E_-+E_+)^2 - (\io)^2} \quad (\text{3D systems}) \label{eqn:higgs_3D}\, ,
\end{eqnarray}
with a slightly different Jacobian for 2D and 3D systems. Limits of integration for $\xi$ are brought to infinity safely since the integrals converge.

\section{The self energy $\Soneloop$}\label{sec:self_energy_oneloop}
We present how the dimensionless part of the self energy is calculated. Denote:
\begin{equation}
    \Soneloop = \Sigma_a + \Sigma_b + \Sigma_c
\end{equation}

\begin{eqnarray}\label{eqn:sigmaa}
    \begin{fmffile}{fmf_appendix_sigmaa}
       \Sigma_a(\iw) =  2\times \parbox{80pt}{
 \begin{fmfgraph*}(80,40)
\fmfset{arrow_len}{7}
    \fmfleft{i1,i2,i3}
    \fmfright{o1,o2,o3}
    \fmf{plain,foreground=(1,,0,,0),tension=0.3}{i2,v1}
    \fmf{phantom,tension=0.4}{i1,v2}
    \fmf{phantom,tension=0.4}{i3,v3}
    \fmf{fermion,tension=0.05}{v1,v2,v3,v1}
\fmf{plain,foreground=(1,,0,,0),tension=0.3,left=1/4}{v3,p2}
\fmf{plain,foreground=(1,,0,,0),tension=0.3,left=1/4}{p3,v2}
    \fmf{fermion,tension=0.05}{p1,p2,p3,p1}
    \fmf{phantom,tension=0.4}{p2,o3}
    \fmf{phantom,tension=0.4}{p3,o1}
    \fmf{plain,foreground=(1,,0,,0),tension=0.3}{p1,o2}
  \end{fmfgraph*}} = 2\int \frac{d\Omega d^d\iq}{(2\pi)^{d+1}} (\Gamma^{(3)}(\iw,\iO,\iq))^2 H_{0,\iq}(\iO) H_{0,\iq}(\io+\iO)
    \end{fmffile}
\end{eqnarray}

\begin{eqnarray}\label{eqn:sigmab}
    \begin{fmffile}{fmf_appendix_sigmab}
            \Sigma_{b}(\io) \,=\,2\times   \parbox{40pt}{
\begin{fmfgraph*}(40,40)
\fmfleft{l}
\fmfright{r}
\fmfset{arrow_len}{7}
\fmfforce{(0w,0.1h)}{l}
\fmfforce{(1w,0.1h)}{r}
\fmfforce{(0.2w,0.2h)}{v1}
\fmfforce{(0.8w,0.2h)}{v2}
\fmfforce{(0.2w,0.8h)}{v4}
\fmfforce{(0.8w,0.8h)}{v3}
\fmf{plain,foreground=(1,,0,,0)}{l,v1}
\fmf{plain,foreground=(1,,0,,0)}{v2,r}
\fmf{fermion,right=0.42}{v1,v2,v3,v4,v1}
\fmf{plain,foreground=(1,,0,,0)}{v3,v4}
\end{fmfgraph*}
} \,=\, 2\int \frac{d\Omega d^d\iq}{(2\pi)^{d+1}} \Gamma^{(4),1}(\iw,\iO,\iq) H_{0,\iq}(\iO)
    \end{fmffile} 
\end{eqnarray}

\begin{eqnarray}\label{eqn:sigmac}
    \begin{fmffile}{fmf_appendix_sigmac}
        \Sigma_{c}(\iw) \,=\, \parbox{40pt}{
\begin{fmfgraph*}(40,40)
\fmfleft{l}
\fmfright{r}
\fmfforce{(-0.1w,0.5h)}{l}
\fmfforce{(1.1w,0.5h)}{r}
\fmfforce{(0.07573w,0.5h)}{v1}
\fmfforce{(0.5w,0.07573h)}{v2}
\fmfforce{(0.92426w,0.5h)}{v3}
\fmfforce{(0.5w,0.92426h)}{v4}
\fmfset{arrow_len}{7}
\fmf{fermion,right=0.42}{v1,v2,v3,v4,v1}
\fmf{plain,foreground=(1,,0,,0)}{l,v1}
\fmf{plain,foreground=(1,,0,,0)}{v3,r}
\fmf{plain,foreground=(1,,0,,0)}{v4,v2}
\end{fmfgraph*} 
} \,=\, \int \frac{d\Omega d^d\iq}{(2\pi)^{d+1}} \Gamma^{(4),2}(\iw,\iO,\iq) H_{0,\iq}(\iO)
    \end{fmffile}
\end{eqnarray}

(Remember that all $\iq$ integrals are limited to the region $\iq<\xi_0^{-1}$).

In all of these diagrams, $\Gamma$ represents the fermionic loops shown in the corresponding Feynman diagrams. $\iw$ is defined to be the \emph{incoming frequency}, and we will evaluate all of the $\Sigma$'s at zero incoming momentum. In this case, the analytical form can be evaluated with Mathematica \cite{Mathematica}. Here we present $\Gamma^{(3)}(\iw,\iO,\iq)$  for completeness. We use simplified notation \(\xi_{--} = \xi_{\ik - \iq}\) and likewise for the \(+\) sign and for \(E_{--/++}\). :
\begin{eqnarray}
   \Gamma^{(3)}(\io,\iO,\iq) &=& -\sum_{\ik,\iv} \Tr[G_{0,\ik}(\iw+\iv)\sigma_1 G_{0,\ik-\iq}(\iv -\iO) \sigma_1 G_{0,\ik}(\iv)\sigma_1]  \nonumber\\ 
&=& -\int \frac{d^d\ik}{(2\pi)^d} \ \Delta_0 \times     \Big( \frac{-\Delta_0^2 - 3E_\ik^2 + (\io)(\iO) + 2E_\ik(\iw - \iO) + \xi^2_\ik + 2\xi_\ik\xi_{--} }{\iw E_\ik(E_{--}^2-(\iO+E_{\ik})^2)(2E_\ik-\iw)}     +  \nonumber\\
&& \frac{\Delta_0^2 + 3E_{--}^2 + \iO(\iO + \iw) - 2E_{--}(2\iO + \iw) - \xi_\ik(\xi_\ik + 2\xi_{--})}{E_{--}((\iO-E_{--})^2 - E_\ik^2)((\iO+\iw - E_{--})^2-E_\ik^2)}  \nonumber \\ 
&& \frac{\Delta_0^2 + 3E_\ik^2 + \iw(\iO+\iw) + 2E_\ik(\iO+2\iw) - \xi_\ik(\xi_\ik + 2 \xi_{--})}{\iw E_\ik(E_{--}^2 - (\iO+\iw + E_\ik)^2)(2E_\ik+\iw)}
\Big)\, . 
\end{eqnarray}



Other fermionic bubbles $\Gamma^{(4),1}$ and $\Gamma^{(4),2}$ can be obtained in the same way. Depending on the dimension, the sum over momentum $\ik$ is replaced by a double integral:
\begin{equation}
    \sum_\ik = N_F\int_{-\infty}^\infty d\xi \times \begin{cases}
        \int_0^{2\pi} d\theta/(2\pi) \quad \text{    2D systems} \\ 
        \int_{-1}^{1} d\cos(\theta)/2 \quad \text{3D systems}
    \end{cases}
\end{equation}

Each of the fermionic bubbles can be solved by numerical integration using the adaptive quadrature algorithm \cite{piessensQuadpack1983} at different $\io,\iO,\iq$.

The next step is to convert the self-energy into dimensionless integrals. Renormalizing all frequencies and momenta: $u = q\xi_0$. For example, $\Sigma^a$ in three dimensions becomes
\begin{eqnarray}
    \Sigma_a(\iw) &=& \int_{-\infty}^{\infty}\frac{d\Omega}{2\pi} \times dK_d \times \int_0^{\sqrt{12}\Delta_0/v_F} {q^2dq}  \ \Gamma^{(3)}(\iw,\iO,\iq))^2 H_{0,\iq}(\iO) H_{0,\iq}(\io+\iO) \label{app:sigmaa}  \\ 
    &=& \int_{-\infty}^{\infty}\frac{d\Omega}{2\pi} \times  \frac{dK_d }{\xi^d} \times \int_0^{1} {u^2du}  \ \frac{1}{\Delta_0^2}(\Gamma^{(3)}(\iw/\Delta_0,\iO/\Delta_0,u))^2 H_{0,u}(\iO/\Delta_0) H_{0,u}(\io/\Delta_0+\iO/\Delta_0) \label{app:sigmab}\\ 
   &=& \frac{\alpha N_F}{\Delta_0
   } \times \underbrace{\int_{-\infty}^\infty \frac{d \Omega / \Delta_0}{2\pi} \int_0^{1}   u^2du\  \Gamma^{(3)}(\iw/\Delta_0,\iO/\Delta_0,u))^2 H_{0,u}(\iO/\Delta_0) H_{0,u}(\io/\Delta_0+\iO/\Delta_0)}_{\displaystyle \large \bar\Sigma_a(\io/\Delta_0)}\label{app:sigmac}
\end{eqnarray}
where the bar on top of $\bar\Sigma$ indicates it is now dimensionless.

\section{Analytic continuation}

\begin{figure}
    \centering
    \includegraphics[width=1\linewidth]{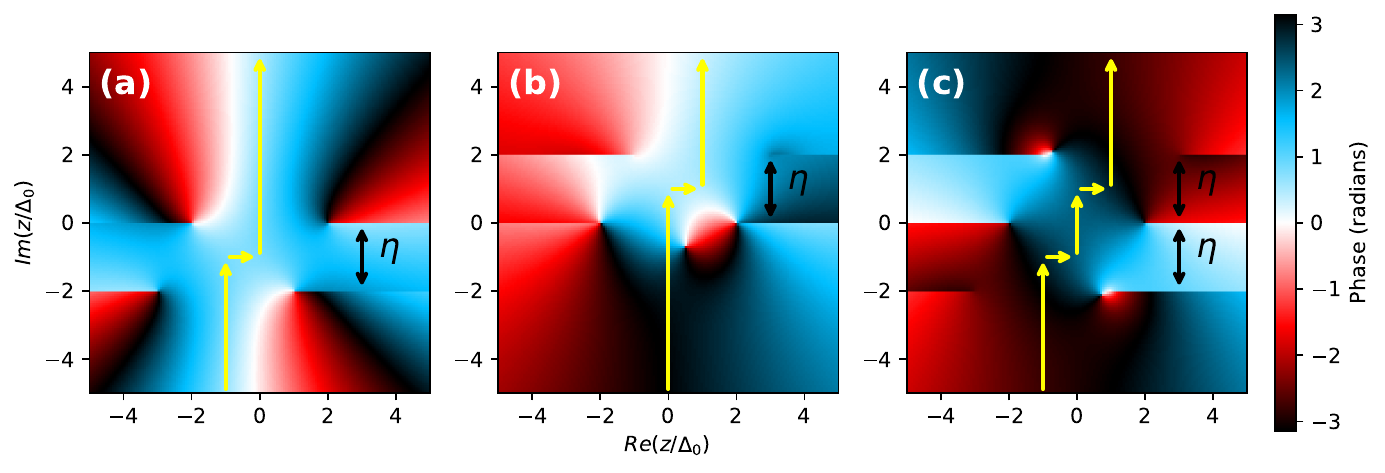}
    \caption{(a), (b), and (c) are the complex phases of the integrands in $\Sigma_a$, $\Sigma_b$, and $\Sigma_c$, respectively. The yellow arrows show the integration contour $\gamma'$ over $\iO \rightarrow z$. They are defined as (a). line segments that connects the points $(-\omega-i\infty) \rightarrow (-\omega-i\eta/2) \rightarrow (-i\eta/2) \rightarrow(i\infty)$. (b) line segments connecting $(-i\infty)\rightarrow(i\eta/2) \rightarrow(\omega + i\eta/2)\rightarrow (\omega+i\infty)$. (c) line segments connecting $(-\omega-i\infty) \rightarrow (-\omega-i\eta/2) \rightarrow (-i\eta/2)\rightarrow(i\eta/2)\rightarrow(\omega+i\eta/2) \rightarrow(\omega+i\infty)$.   }
    \label{fig:sup_contour}
\end{figure}

Integrands in Eq.~(\ref{app:sigmaa},\ref{app:sigmab},\ref{app:sigmac}) are not analytical in parts of the complex plane. For example, let $\iO \rightarrow z$ with $z$ being a complex number, the self energy $\Sigma_a$ becomes:
\begin{equation}
    \Sigma_{a}(\iw) = -2i\int_\gamma \frac{dz d^d\iq}{(2\pi)^{d+1}} (\Gamma^{(3)}(\iw,z,\iq))^2 H_{0,\iq}(z) H_{0,\iq}(\io+z)\, , 
\end{equation}
where $\gamma$ is the integration along the imaginary axis (from $-i\infty$ to $i\infty$). Analytic continuation will shift the incoming frequency to $\io \rightarrow \omega +i\eta$. In other words, we move $\io$ on the complex plane such that:
\begin{equation} \label{eqn:sup_analytical_continuation_wrong}
    \int_\gamma H(z+\io) \dots \  dz \xleftrightarrow{\text{Analytic continuation}} \int_{\gamma} H(z+\omega+i\eta) \dots \ dz\, .
\end{equation}

In this example, $H(z+\iw)$ has branch cuts $z \in  [2\Delta_0-\iw,\infty-\iw) \cup (-\infty-\iw,-2\Delta_0-\iw]$ and analytically continuing $\io \rightarrow \omega+i\eta$ shifts the branch cuts to be $z \in  [2\Delta_0-\omega - i\eta,\infty-i\eta) \cup (-\infty-i\eta,-2\Delta_0-\omega-i\eta]$. Since the contour $\gamma$ lies on the imaginary axis, the branch cuts of the analytically continued result will cross the contour $\gamma$ for $\omega>2\Delta_0$. In this case, the left- and right-hand sides of Eq.~(\ref{eqn:sup_analytical_continuation_wrong}) are no longer smoothly connected, and analytic continuation fails. The solution to this problem is to change the right-hand side of Eq.~(\ref{eqn:sup_analytical_continuation_wrong}) to 
\begin{equation} \label{eqn:sup_analytical_continuation_correct}
    \int_\gamma H(z+\io) \dots \  dz \xleftrightarrow{\text{Analytic continuation}} \int_{\gamma'} H(z+\omega+i\eta) \dots \ dz\, , 
\end{equation}
i.e., change the integration contour of the right-hand side from $\gamma$ to $\gamma'$. $\gamma'$ is chosen such that it never crosses any branch cut or singularity. We pick the contours shown in Fig.~(\ref{fig:sup_contour}). These choices of $\gamma'$ guarantee that the right-hand side of Eq.~(\ref{eqn:sup_analytical_continuation_correct}) can be smoothly deformed into the left-hand side via shifting both the contour $\gamma'\rightarrow \gamma$ and $\omega+i\eta \rightarrow i\omega$. We have verified that this method gives the correct result for analytically solvable cases. Practically, this means that we are allowed to \emph{first perform the analytic continuation} $i\omega \rightarrow \omega+i\eta$ and then integrate numerically with the contour $\gamma'$ to evaluate the self energies Eq.~(\ref{eqn:sigmaa},\ref{eqn:sigmab},\ref{eqn:sigmac}). 

\begin{figure}[!h]
    \centering
    \includegraphics[width=1\linewidth]{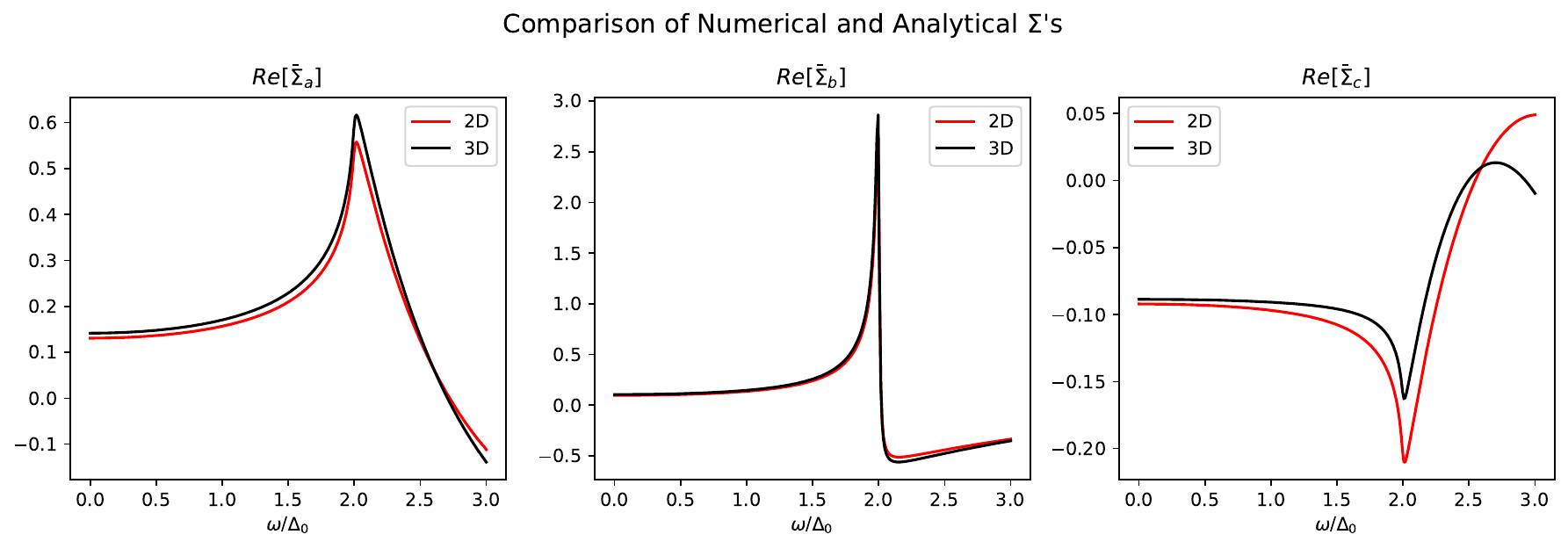}
    \caption{Real parts for the self energies. $\eta = 0.01\Delta_0$}
    \label{fig:Re_sigmas}
\end{figure}

\begin{figure}[!h]
    \centering
    \includegraphics[width=1\linewidth]{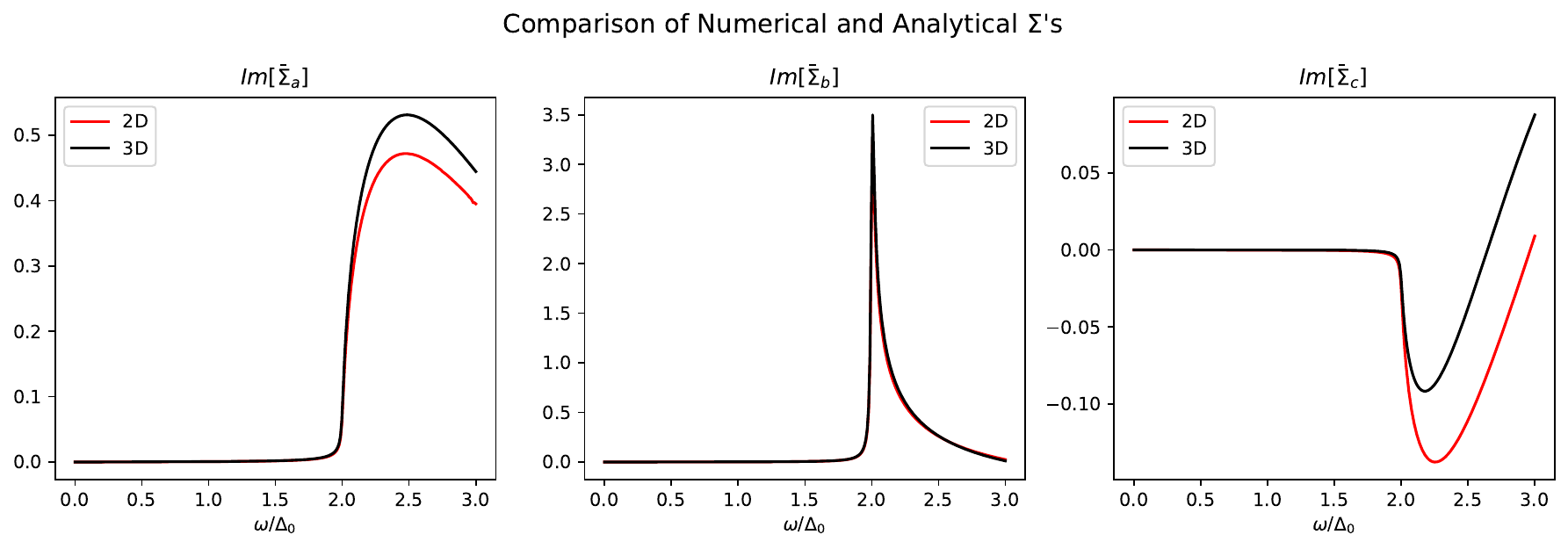}
    \caption{Imaginary parts for the self energies. $\eta = 0.01\Delta_0$}
    \label{fig:Im_sigmas}
\end{figure}

The final results of each of the renormalised self energies ($\bar{\Sigma}_a$,$\bar{\Sigma}_b$,$\bar{\Sigma}_c$) are presented in Fig.~(\ref{fig:Re_sigmas},\ref{fig:Im_sigmas}) for completeness.
\newpage

\bibliography{ref_sup}

%% file: feyn_macro.tex
\begin{enumerate}
    \item[1.] The bare Nambu propagator: 
\end{enumerate}

\begin{equation}\label{eqn:feynrule_1}
\begin{fmffile}{electron_greens_function}
\parbox{40pt}{
\begin{fmfgraph*}(40,40)
\fmfset{arrow_len}{7}
    \fmfleft{i}
    \fmfright{j}
    \fmf{fermion}{i,j}
\end{fmfgraph*}}
\end{fmffile}  \quad  = \quad  G_{0,\ik}(\iv) = \frac{\iv + \xi_\ik \sigma_3 + \Delta_0 \sigma_1}{ E^2_\ik- (\iv)^2 } 
\end{equation}

\begin{enumerate}
    \item[2.] The bare propagator of the \(h_{i}\) field:
\end{enumerate}

\begin{equation}\label{eqn:feynrule_2}
\begin{fmffile}{h_bare_greens_function}
\parbox{40pt}{
\begin{fmfgraph*}(40,40)
\fmfset{arrow_len}{7}
    \fmfleft{i}
    \fmfright{j}
    \fmf{dashes,foreground=(1,,0,,0)}{i,j}
\end{fmfgraph*}}
\end{fmffile}  \quad  = \quad  \frac{V}{2}
\end{equation}  

\begin{enumerate}
    \item[3.] The interaction vertex between the $h$ and the $\Psi$ field:
\end{enumerate}

\begin{equation}\label{eqn:feynrule_3}
\begin{fmffile}{interaction_h_psi}
\parbox{40pt}{
\begin{fmfgraph*}(40,40)
\fmfset{arrow_len}{7}
    \fmfleft{i}
    \fmfright{j1,j2}
    \fmf{dashes,foreground=(1,,0,,0)}{i,v}
    \fmf{fermion}{j1,v,j2}
\end{fmfgraph*}}
\end{fmffile}  \quad  = \quad  -\sigma_1
\end{equation}  

\begin{enumerate}
    \item[4.] A term contributing to the vacuum expectation value of the \(h\) field:
\end{enumerate}
\begin{equation}\label{eqn:feynrule_4}
\begin{fmffile}{h_terminate_term}
\parbox{40pt}{
\begin{fmfgraph*}(40,40)
\fmfset{arrow_len}{7}
    \fmfleft{i}
    \fmfright{j}
    \fmf{dashes,foreground=(1,,0,,0)}{i,j}
    \fmfv{decor.shape=cross,decor.filled=empty,decor.size=10}{j}
\end{fmfgraph*}}
\end{fmffile}  \quad  = \quad  -\frac{2\Delta_0}{V}
\end{equation}

%% file: LW_macro.tex
\begin{eqnarray}\label{eqn:LWfunctional}
    \Phi[G,\textcolor{red}{H}] \,  = \,  \begin{fmffile}{LW-functional_Hartree}
        \frac{1}{2}\parbox{30pt}{  \begin{fmfgraph*}(30,20)
            \fmfset{arrow_len}{6}
            \fmfset{arrow_ang}{20}  
            \fmftop{t}
            \fmfbottom{b}
            \fmf{heavy,left}{b,b}
            \fmf{heavy,left}{t,t}
            \fmf{plain,foreground=(1,,0,,0)}{b,t}
        \end{fmfgraph*}}
        + \parbox{30pt}{  \begin{fmfgraph*}(30,20)
            \fmfset{arrow_len}{6}
            \fmfset{arrow_ang}{20}  
            \fmftop{t}
            \fmfbottom{b}
            \fmf{heavy,left}{t,t}
            \fmf{plain,foreground=(1,,0,,0)}{b,t}
            \fmfv{decor.shape=cross,decor.filled=empty,decor.size=10}{b}
        \end{fmfgraph*}}
    \end{fmffile}  
    +\begin{fmffile}{LW-functional11}
\frac{1}{2}\,\parbox{40pt}{
\begin{fmfgraph*}(40,40)
    \fmfsurround{v1,v2}
    \fmfset{arrow_len}{6}
    \fmfset{arrow_ang}{20}  
    \fmf{heavy,right}{v1,v2,v1}
    \fmf{plain,foreground=(1,,0,,0)}{v1,v2}
\end{fmfgraph*}
} \, 
\,+\,\frac{1}{4}\,
\parbox{40pt}{
\begin{fmfgraph*}(40,40)
\fmfsurround{v1,v2,v3,v4}
            \fmfset{arrow_len}{6}
            \fmfset{arrow_ang}{20}  
\fmf{heavy,right=0.42}{v1,v2,v3,v4,v1}
\fmf{plain,foreground=(1,,0,,0)}{v1,v3}
\fmf{plain,foreground=(1,,0,,0)}{v4,v2}
\end{fmfgraph*}
}
\end{fmffile}
\begin{fmffile}{LW-functional22}
+\frac{1}{3!}\, \quad \quad  \parbox{40pt}{
\begin{fmfgraph*}(25,40)
            \fmfset{arrow_len}{6}
            \fmfset{arrow_ang}{20}  
\fmfstraight
\fmfleftn{l}{5}
\fmfrightn{r}{5}
\fmf{heavy,right}{l5,l1}
\fmf{heavy}{l1,l2,l3,l4,l5}
\fmf{heavy,right}{r1,r5}
\fmf{heavy}{r5,r4,r3,r2,r1}
\fmf{plain,foreground=(1,,0,,0)}{l2,r2}
\fmf{plain,foreground=(1,,0,,0)}{l3,r3}
\fmf{plain,foreground=(1,,0,,0)}{l4,r4}
\end{fmfgraph*}}
\quad \, + \frac{1}{3!}\, \quad \quad  \parbox{40pt}{
\begin{fmfgraph*}(25,40)
            \fmfset{arrow_len}{6}
            \fmfset{arrow_ang}{20}  
\fmfstraight
\fmfleftn{l}{5}
\fmfrightn{r}{5}
\fmf{heavy,right}{l5,l1}
\fmf{heavy}{l1,l2,l3,l4,l5}
\fmf{heavy,right}{r1,r5}
\fmf{heavy}{r5,r4,r3,r2,r1}
\fmf{plain,foreground=(1,,0,,0)}{l2,r2}
\fmf{plain,foreground=(1,,0,,0)}{l3,r4}
\fmf{plain,foreground=(1,,0,,0)}{l4,r3}
\end{fmfgraph*}
}
\end{fmffile}
\end{eqnarray}

%% file: Dyson_G.tex
\Pi[G,\textcolor{red}{H}] = 
\begin{fmffile}{eqn_selfenergy_PI}
    \parbox{30pt}{\begin{fmfgraph*}(30,20)
        \fmfset{arrow_len}{6}
        \fmfset{arrow_ang}{20}  
        \fmftop{t}
        \fmfbottom{b}
        \fmf{heavy,left}{t,t}
        \fmf{plain,foreground=(1,,0,,0)}{b,t}
    \end{fmfgraph*}} + \parbox{30pt}{
    \begin{fmfgraph*}(30,20)
        \fmfset{arrow_len}{6}
        \fmfset{arrow_ang}{20}  
        \fmftop{t}
        \fmfbottom{b}
        \fmfv{decor.shape=cross,decor.filled=empty,decor.size=10}{t}
        \fmf{plain,foreground=(1,,0,,0)}{b,t}
    \end{fmfgraph*}} \, + \,  \parbox{40pt}{
    \begin{fmfgraph*}(40,40)
        \fmfset{arrow_len}{6}
        \fmfset{arrow_ang}{20}  
        \fmfleft{i}
        \fmfright{f}
        \fmf{heavy}{i,f}
        \fmf{plain,foreground=(1,,0,,0),left}{i,f}
    \end{fmfgraph*}
    } \, \, +\,\,  
    \parbox{30pt}{\begin{fmfgraph*}(30,30) 
        \fmfset{arrow_len}{6}
        \fmfset{arrow_ang}{20}  
        \fmfstraight
        \fmfbottom{b1,b2}
        \fmftop{t1,t2}
        \fmf{heavy}{b1,t1,t2,b2}
        \fmf{plain,foreground=(1,,0,,0)}{b1,t2}
        \fmf{plain,foreground=(1,,0,,0)}{b2,t1}
    \end{fmfgraph*}}
    \,\, + \, \, \parbox{40pt}{
    \begin{fmfgraph*}(40,20)
        \fmfset{arrow_len}{6}
        \fmfset{arrow_ang}{20}  
        \fmfstraight
        \fmftop{t1,t2,t3}
        \fmfbottom{b1,b2,b3}
        \fmf{heavy}{b1,b2,b3}
        \fmf{heavy}{t1,t2,t3}
        \fmf{heavy,right}{t3,t1}
        \fmf{plain,foreground=(1,,0,,0)}{b1,t1}
        \fmf{plain,foreground=(1,,0,,0)}{b2,t2}
        \fmf{plain,foreground=(1,,0,,0)}{b3,t3}
    \end{fmfgraph*}
    } \, \, + \,\, 
    \parbox{40pt}{
            \begin{fmfgraph*}(40,20)
        \fmfset{arrow_len}{6}
        \fmfset{arrow_ang}{20}  
        \fmfstraight
        \fmftop{t1,t2,t3}
        \fmfbottom{b1,b2,b3}
        \fmf{heavy}{b1,b2,b3}
        \fmf{heavy}{t1,t2,t3}
        \fmf{heavy,right}{t3,t1}
        \fmf{plain,foreground=(1,,0,,0)}{b1,t2}
        \fmf{plain,foreground=(1,,0,,0)}{b2,t1}
        \fmf{plain,foreground=(1,,0,,0)}{b3,t3}
    \end{fmfgraph*}
    }
\end{fmffile}

%% file: Dyson_H.tex
\Sigma[G,\textcolor{red}{H}] \, \, = \, \, 
\begin{fmffile}{self_energy_sigma}
1\,
\parbox{40pt}{
\begin{fmfgraph*}(40,40)
\fmfset{arrow_len}{6}
\fmfset{arrow_ang}{20}  
\fmfleft{l}
\fmfright{r}
\fmfset{arrow_len}{7}
\fmfforce{(0w,0.5h)}{l}
\fmfforce{(1w,0.5h)}{r}
\fmfforce{(0.15w,0.5h)}{v1}
\fmfforce{(0.85w,0.5h)}{v2}
\fmf{plain,foreground=(1,,0,,0)}{l,v1}
\fmf{plain,foreground=(1,,0,,0)}{v2,r}
\fmf{heavy,right=1}{v1,v2,v1}
\end{fmfgraph*}
}
\,\,
+\,1 \quad
\parbox{40pt}{
\begin{fmfgraph*}(40,40)
\fmfset{arrow_len}{6}
\fmfset{arrow_ang}{20}  
\fmfleft{l}
\fmfright{r}
\fmfforce{(-0.1w,0.5h)}{l}
\fmfforce{(1.1w,0.5h)}{r}
\fmfforce{(0.07573w,0.5h)}{v1}
\fmfforce{(0.5w,0.07573h)}{v2}
\fmfforce{(0.92426w,0.5h)}{v3}
\fmfforce{(0.5w,0.92426h)}{v4}
\fmfset{arrow_len}{7}
\fmf{heavy,right=0.42}{v1,v2,v3,v4,v1}
\fmf{plain,foreground=(1,,0,,0)}{l,v1}
\fmf{plain,foreground=(1,,0,,0)}{v3,r}
\fmf{plain,foreground=(1,,0,,0)}{v4,v2}
\end{fmfgraph*}
}
\quad
+2\,  \parbox{80pt}{
\begin{fmfgraph*}(80,40)
\fmfset{arrow_len}{6}
\fmfset{arrow_ang}{20}  
\fmfleft{i1,i2,i3}
\fmfright{o1,o2,o3}
\fmf{plain,foreground=(1,,0,,0),tension=0.3}{i2,v1}
\fmf{phantom,tension=0.4}{i1,v2}
\fmf{phantom,tension=0.4}{i3,v3}
\fmf{heavy,tension=0.05}{v1,v2,v3,v1}
\fmf{plain,foreground=(1,,0,,0),tension=0.3,left=1/4}{v3,p2}
\fmf{plain,foreground=(1,,0,,0),tension=0.3,left=1/4}{p3,v2}
\fmf{heavy,tension=0.05}{p1,p2,p3,p1}
\fmf{phantom,tension=0.4}{p2,o3}
\fmf{phantom,tension=0.4}{p3,o1}
\fmf{plain,foreground=(1,,0,,0),tension=0.3}{p1,o2}
\end{fmfgraph*}
}
\end{fmffile}\, . 

%% file: Gap_equation_Full.tex
\begin{equation}\label{eqn:gap_equation_FullGF}
    \begin{fmffile}{fmf_gap_equation_full_GF}
        \parbox{40pt}{\begin{fmfgraph*}(40,30)
            \fmfset{arrow_len}{6}
            \fmfset{arrow_ang}{20}  
            \fmftop{t}
            \fmfbottom{b}
            \fmf{heavy,left}{t,t}
            \fmf{plain,foreground=(1,,0,,0)}{b,t}
        \end{fmfgraph*}} + \parbox{40pt}{
        \begin{fmfgraph*}(40,30)
            \fmfset{arrow_len}{6}
            \fmfset{arrow_ang}{20}  
            \fmftop{t}
            \fmfbottom{b}
            \fmfv{decor.shape=cross,decor.filled=empty,decor.size=10}{t}
            \fmf{plain,foreground=(1,,0,,0)}{b,t}
        \end{fmfgraph*}} = 0 
    \end{fmffile} \, . 
\end{equation}